\documentclass[11pt]{article}
\usepackage{amssymb,amsmath}
\usepackage{bm} 
\usepackage{booktabs} 
\usepackage{array}
\usepackage{latexsym}
\usepackage{graphicx}
\usepackage{color}
\usepackage{datetime}
\usepackage[nosort]{cite}
\usepackage{verbatim}
\usepackage{enumerate}
\usepackage{chngpage} 
\usepackage{mathrsfs}
\usepackage{euscript}
\usepackage{psfrag}

\usepackage{mciteplus}

\usepackage[colorlinks=true,      linkcolor=blue,      urlcolor=blue,
            filecolor=blue,      citecolor=blue,       pdfstartview=FitH,
                        pdfpagemode=UseNone,      bookmarksopen=true]{hyperref}  
\usepackage[all]{hypcap}     

\usepackage{tensor}
\usepackage{physics}
\usepackage{tikz}

\renewcommand{\comm}[1]{} 

\def\({\left(}
\def\){\right)}
\def\[{\left[}
\def\]{\right]}

\def\coeff#1#2{{\textstyle \frac{#1}{#2}}}

\def\One{{\hbox{ 1\kern-.8mm l}}}

\def\barray{\begin{array}}
\def\earray{\end{array}}
\def\be{\begin{equation}}
\def\ee{\end{equation}}
\def\bea{\begin{eqnarray}}
\def\eea{\end{eqnarray}}
\def\bal{\begin{align}}
\def\eal{\end{align}}

\numberwithin{equation}{section} 

\definecolor{cardinal}{rgb}{0.6,0,0}
\definecolor{darkgreen}{rgb}{0,0.4,0}
\definecolor{golden}{rgb}{0.92, 0.7, 0}
\definecolor{midnight}{rgb}{0, 0, 0.5}
\definecolor{darkblue}{rgb}{0, 0, 0.7}
\definecolor{darkred}{rgb}{0.6, 0, 0}
\definecolor{purple}{rgb}{0.5, 0, 0.5}

\def\IR{\mathbb{R}}

\def\IT{\mathbb{T}}

\def\ZZ{\mathbb{Z}}

\def\cA{{\cal A}}
\def\cB{{\cal B}}

\def\cF{{\cal F}}

\def\cI{{\cal I}}

\def\cK{{\cal K}}

\def\cN{{\cal N}}
\def\cO{{\cal O}}
\def\cP{{\cal P}}

\newcommand{\stP}{\mathcal{P}}

\begin{document}

\phantom{AAA}
\vspace{-10mm}

\begin{flushright}
%
%
\end{flushright}

\vspace{1.9cm}

\begin{center}

{\huge {\bf Delaying the Inevitable: Tidal Disruption}}\\
{\huge {\bf \vspace*{.25cm} in Microstate Geometries }}

\vspace{1cm}

{\large{\bf {Iosif Bena$^1$,~Anthony Houppe$^{1}$   and  Nicholas P. Warner$^{1,2,3}$}}}

\vspace{1cm}

$^1$Institut de Physique Th\'eorique, \\
Universit\'e Paris Saclay, CEA, CNRS,\\
Orme des Merisiers, Gif sur Yvette, 91191 CEDEX, France \\[12 pt]
\centerline{$^2$Department of Physics and Astronomy}
\centerline{and $^3$Department of Mathematics,}
\centerline{University of Southern California,} 
\centerline{Los Angeles, CA 90089, USA}

\vspace{10mm} 
{\footnotesize\upshape\ttfamily iosif.bena @ ipht.fr, anthony.houppe @ ipht.fr, warner @ usc.edu} \\

\vspace{2.2cm}
 
\textsc{Abstract}

\end{center}

\begin{adjustwidth}{3mm}{3mm} 
 
\vspace{-1.2mm}
\noindent
Microstate geometries in string theory replace the black-hole horizon with a smooth geometric ``cap'' at the horizon scale. In geometries constructed using superstratum technology, this cap has the somewhat surprising property  that induces very large tidal deformations on infalling observers that are far away from it.  We find that this large-distance amplification of the tidal effects is also present in horizonless microstate geometries constructed as bubbling solutions, but can be tamed by suitably arranging the bubbles to reduce the strength of some of the gravitational multipole moments. However, despite this taming, these tidal effects still become large at a significant distance from the microstructure.  This result suggests that an observer will not fall unharmed into the structure replacing the black hole horizon.

%
%
\end{adjustwidth}

\thispagestyle{empty}
\newpage


\baselineskip=17pt
\parskip=5pt

\setcounter{tocdepth}{2}
\tableofcontents

\baselineskip=15pt
\parskip=3pt

\newpage

\section{Introduction}
\label{sec:Intro}

Microstate geometries are, by definition, smooth, solutions of supergravity that have the same mass and charges as a black hole and resemble the black-hole solution at all scales larger than the horizon scale. However, instead of having a horizon and singularity, microstate geometries ``cap off'' smoothly at large red-shifts (see, for example, \cite{Bena:2006kb, Bena:2007qc, Bena:2016ypk, Bena:2017xbt, Heidmann:2017cxt, Bena:2017fvm, Avila:2017pwi, Ceplak:2018pws,Heidmann:2019zws, Heidmann:2019xrd,Warner:2019jll}).   To date,  huge families of such geometries have been constructed and they differ from one another only at the scale of the cap.  Such geometries provide  explicit, well-controlled support for the kind of horizon-scale structure that is needed to solve the black hole information paradox \cite{Mathur:2009hf, Almheiri:2012rt}.  

If a horizonless structure replaces the black hole horizon, it is very important to see how and where the deviations from black-hole behavior become significant. Until recently it was thought that the scrambling of an incoming probe into the black hole microstate geometry would happen when the probe encounters detailed microstructure at the bottom (or cap) of the geometry.  It has also been suggested that, even on scales where classical physics could naively be trusted, the presence of a large number of microstates enhances the quantum tunneling, and hence the absorption of matter by a microstate geometry should be an intrinsically quantum phenomenon \cite{Kraus:2015zda,Bena:2015dpt}.  While these ideas may be important in the scrambling process,  recent work \cite{Tyukov:2017uig,Bena:2018mpb} has suggested that a more mundane mechanism may also be a leading effect:  tidal forces of an infalling particle can  become large, of order the Planck/Compactification scale, when the particle is a long distance away from the cap.


In this paper we will examine tidal stresses in a broader class of microstate geometries, and show that, while large tidal forces will typically arise when the probe is at a long distance from the cap, the region of the onset of these forces can be pushed closer to the cap carefully adjusting the multipole moments of the metric in the microstate geometry.

For classical black holes, the simplest estimate of tidal forces is obtained from the Kretschmann scalar (Riemann squared):
\begin{equation} 
\cI ~=~ R^{\mu \nu \rho \sigma} \, R_{\mu \nu \rho \sigma} \,.
\end{equation} 
This has dimensions of {\it length}$^{-4}$ and in a black-hole solution it is proportional to the square of the mass, $m^2$.  Thus
\begin{equation} 
\cI ~\sim~ \frac{m^2}{r^6}  \,,
\end{equation} 
where $r$ is the radial coordinate.  The horizon is located at $r \sim m$, and so 
\begin{equation} 
\cI_{\rm horizon}  ~\sim~ \frac{1}{m^4}  \,,
\end{equation} 
which indicates that tidal forces on observers crossing the horizon become extremely small for large black holes. A more sophisticated analysis using geodesic deviation for infalling observers confirms this expectation.  

In this paper we will compute the tidal forces in microstate geometries of the supersymmetric three-charge black hole in five dimensions. In the duality frame in which the charges of this black hole correspond to $N_1$ D1 branes, $N_5$ D5 branes and $N_P$ quanta of momentum, the throat of the black-hole solution is a fibration of a deformed three-sphere over the extremal BTZ geometry.  Since the BTZ geometry is AdS$_3$  divided by a discrete group, it has constant curvature.  The infalling geodesics are then trivial to compute and the tidal stress\footnote{This is proportional to the Riemann tensor, and has the units of an acceleration per unit length, which means  {\it length}$^{_2}$. } is of order $R_{AdS}^{-2}$.

The  microstate geometries we consider have the same mass and charges as the black hole, and the same asymptotic AdS$_3 \times$ S$^3 \times \IT^4$ (or K$3$) region. Thus they are dual to pure states of the $(1+1)$-dimensional CFT with central charge $6 N_1 N_5$ that counts the entropy of the black hole. 

There are two standard approaches to constructing such microstate geometries:  (i) multi-centered bubbling microstate geometries and (ii) superstrata. Superstrata have the advantage that their holographically-dual CFT states are well understood at the orbifold point \cite{Giusto:2015dfa,Galliani:2016cai,Ceplak:2018pws}. However, they are only smooth in the D1-D5-P duality frame and in a few other duality frames in which the momentum charge along the common D1-D5 direction remains a momentum charge \footnote{These frames are obtained by T-dualities along the $\IT^4$ and S-dualities, which are the duality transformations that preserve the AdS$_3 \times$ S$^3$ region of the microstate geometries.}. In contrast, the bubbling microstate geometries are smooth in all possible duality frames, but their AdS-CFT dictionary in the D1-D5-P frame is not understood.

One would naively expect that the deviations caused by the microstate structure would appear as multipole corrections to the BTZ geometry, which are only important very close to this structure. However, in superstrata it was found that for an infalling observer these multipole corrections give rise to large tidal forces ``half-way down''  the  throat \cite{Tyukov:2017uig,Bena:2018mpb}\footnote{That is, they become large when the radial coordinate, $r$, is at the geometric mean of the scales at the top and bottom of the capped-BTZ throat.}. 

In the D1-D5-P duality frame there are two reasons for this rather unexpected large-distance amplification of the effects of the microstructure. The first has to do with the effect of the momentum charge of the D1-D5-P black hole (we will explain this in more detail in a moment).  The second is the ultra-relativistic boost of the probe as it falls into the geometry; this boost magnifies the curvature deviations\footnote{It is also possible to compute the tidal force in five-dimensional microstate geometries, where the three charges of the black hole correspond to M2 branes wrapping three orthogonal two-tori inside a  $\IT^6$, and in these solutions the large-distance amplification of the tidal distortion comes only because of the boost of the infalling probe.}.

The major focus of our paper will be the calculation of tidal forces in bubbling multi-centered microstate geometries. Since these geometries are smooth in all duality frames, this calculation can be done both in the D1-D5-P duality frame, where it can be compared to the superstratum tidal-force calculation   \cite{Tyukov:2017uig,Bena:2018mpb}, and also in other duality frames in which one cannot construct smooth superstratum solutions. 
What makes multi-centered geometries more interesting than superstrata is that their construction allows one to  control the multipole moments of the gravitational field and one can use this to soften the tidal impact and create a more black-hole-like effect on infalling probes. 

In the D1-D5-P duality frame, both superstrata and  multi-centered microstate geometries have three important length scales: $R_{AdS} = (Q_1 Q_5)^{\frac{1}{4}}$, $b \sim \sqrt{Q_P}$, \footnote{The charges $Q_1,Q_5$ and $Q_P$ are ``supergravity'' charges with dimension  {\it length}$^{-2}$.  They are related to the dimensionless, quantized D1, D5 and P charges in \eqref{QstoNs}.} and a parameter, $a$, which corresponds to the ``size'' of the microstructure in the un-warped $\IR^3$ base space on which both types of solutions are constructed. In superstrata and in generic bubbling microstate geometries with a long throat, $a$ is also proportional to the left-moving angular momentum, $J_L$. As $a$ decreases, the BTZ throat of the microstate geometry becomes longer and longer, and the geometry resembles more and more the classical black-hole geometry. However, because of the warping, the physical size of the microstructure is independent of $a$ \cite{Bena:2006kb,Bena:2016ypk}.

We will work in the regime $Q_P,\, (Q_1 Q_5)^{\frac{1}{2}}\, \gg\, a^2 $, in which the three-dimensional part of the microstate geometries have a very long capped BTZ throat and hence resemble the BTZ solution to arbitrary precision. The metric has three distinct regions:
\begin{itemize} 
\item  $r \gtrsim \sqrt{Q_P}$:  The AdS$_3$ region, or the upper region of the BTZ geometry.
\item $a \ll r \ll \sqrt{Q_P}$:  The AdS$_2 \times S^1$ throat of the BTZ geometry.
\item  $r \lesssim a$:  ``The cap,''  in which the BTZ throat is smoothly rounded off, usually at high red-shift.
\end{itemize} 
%


The tidal forces on an infalling particle in the capped-BTZ throat of superstrata was found in  \cite{Tyukov:2017uig,Bena:2018mpb} to be of the form
\begin{equation}
\frac{ a^2 \, Q_P \,E^2 }{r^{6}} \,,
  \label{TidalDom1}
\end{equation}
where $E$ is the energy (per unit rest-mass) of the infalling geodesic. For a particle released from the top of the BTZ throat, one has $E^2 \sim \frac{\sqrt{Q_P}}{R_y}$, where $R_y$ is the radius of the common D1-D5 circle. It is this tidal force that comes to dominate over the standard BTZ curvature in the middle of the throat, high above the microstructure. 

It was shown in \cite{Bena:2018mpb} that the term (\ref{TidalDom1}) is a universal feature of the tidal force felt by an infalling observer in a superstratum geometry. The factor of $a^2$ comes from the (small) left- moving angular momentum of the superstratum solution, which cannot be set to zero without creating an AdS$_2$ throat of infinite depth, thereby decoupling the cap  from the asymptotic region \cite{Bena:2018bbd}. This raises the question as to whether one might be able to cancel the large tidal deformations by using certain bubbling microstate geometries in which $J_L$ can be set to zero while keeping the length of the capped-BTZ throat finite \cite{Bena:2006kb}. The purpose of this paper is to show that terms of the form  (\ref{TidalDom1}) can indeed be canceled, but this only delays the onset of large tidal forces.  Higher multipole moments are non-zero and these lead to tidal disruption further down the throat. 

In Section \ref{Sect:tides} we give a brief review of the essentials of tidal forces and then, in Section \ref{Sect:SStides} we give a brief review the results of  \cite{Tyukov:2017uig,Bena:2018mpb} on tidal forces on probes falling into superstrata. In Section \ref{sec:multi-centered}, we consider the tidal force in five-centered microstate geometries in which the solution can be chosen to have a $\ZZ_2$ symmetry that causes the angular momentum, $J_L$, to vanish identically.  We show that when $J_L$ vanishes, the coefficient of    (\ref{TidalDom1}) vanishes as well.  We then look for the next sub-leading tidal terms and see that they still create large tidal forces, but do so deeper in the throat.  Since the large tidal forces arise at a large distance away from the cap, the result should not depend on the detailed distribution of charge sources.  We can therefore follow the philosophy in \cite{Bena:2018mpb}, and compute the tidal forces more simply by replacing the topological bubbles of the microstate geometry by black rings that localize in the cap.  We describe this in Section \ref{sec:BlackenedBubbles}.   The tidal computations turn out to be extremely demanding at the computational level and we found it essential to streamline them by computing the six-dimensional Riemann tensor analytically. This useful expression may be found in the Appendix.  Out final comments appear in   Section \ref{sec:Conclusions}.

\section{Tides}
\label{Sect:tides}

When one refers to tidal forces, one starts with an observer following a time-like geodesic through the geometry.  If this 
 geodesic has a proper velocity, $V^\mu = \frac{dx^\mu}{d \tau}$, then the equation of geodesic deviation is:
\begin{equation}
A^\mu ~\equiv~ \frac{D^2 S^\mu}{d \tau^2}  ~=~ - {R^\mu}_{\nu \rho \sigma} \, V^\nu S^\rho   V^\sigma \,,
  \label{Geodev1}
\end{equation}
where $S^\rho$ is the deviation vector.  By synchronizing the proper time of neighboring geodesics, one can arrange $S^\rho V_\rho = 0$ over the family of geodesics.  Thus $S^\rho$ is a space-like vector in the rest-frame of the geodesic observer.  One can re-scale $S^\mu$ at any one point so that $S^\mu S_\mu =1 $ and then $A^\mu$ represents the acceleration per unit distance, or the tidal stress.  The skew-symmetry of the Riemann tensor means that $ A^\mu V_\mu  =0$, and so the tidal acceleration is similarly space-like, representing the tidal stress in the rest-frame of the infalling observer with velocity, $V^\mu$.

It is convenient to define the ``tidal tensor'' along the geodesic
\begin{equation}
{\cA^\mu}_\rho ~\equiv~ - {R^\mu}_{\nu \rho \sigma} \, V^\nu \, V^\sigma \,,
  \label{cAdefn}
\end{equation}
The symmetries of the Riemann tensor imply that $\cA_{\mu \nu} = \cA_{\nu \mu}$ and $\cA_{\mu \nu} V^\nu =0$.  It follows that $\cA_{\mu \nu}$ is diagonalizable and its non-trivial eigenvectors are space-like.  The norm of the tidal tensor 
\begin{equation}
|\cA| ~\equiv~  \sqrt{{\cA^\mu}_\rho\, {\cA^\rho}_\mu}  \,.
  \label{cAnorm}
\end{equation}
therefore provides an excellent measure of (and bound upon)  the tidal forces experienced by the geodesic observer.  Indeed, the maximal tidal stress is bounded between $\frac{1}{\sqrt{s}}  |\cA|$ and $|\cA|$, where $s$ is the number of spatial directions. 

This paper will focus on calculating $|\cA|$ in a variety of microstate geometries.  It is also useful to note that since $V^\mu = \frac{dx^\mu}{d \tau}$ is dimensionless, $\cA$ has the same dimensions as the curvature tensor,  {\it length}$^{-2}$.

The geometries we will consider are all asymptotic to AdS$_3$ $\times S^3$, and we will choose geodesics that start from rest in the asymptotically-AdS region and penetrate deep into the interior.  This will mean choosing geodesics with vanishing angular momenta on the $S^3$, and with no momentum along the D1-D5 common circle, so that there are no angular momentum barriers. 

Since all the metrics we consider are BPS, the geometries are time independent, and so the geodesics will have a conserved energy, $E$.  The energy $E$ will  be determined by the release point of the probe geodesic in the asymptotically AdS region.  This will then typically leave the ``radial infall'' to be determined through the conserved quantity obtained from the metric:
\begin{equation}
g_{\mu \nu}\,\frac{dx^\mu}{d \tau} \,\frac{dx^\nu}{d \tau} ~=~ -1  \,.
  \label{metcons}
\end{equation}
We will compute $|\cA|$ for these geodesics.

It was evident from the work of \cite{Tyukov:2017uig,Bena:2018mpb} that the ultra-relativistic boost, created by infall from the asymptotic region, significantly enhances the tidal forces on probes. Intuitively this is the same as hitting rough road at excessive speed.  Since  $|\cA|$ is quadratic in velocities, it has terms that are quadratic in $E$, and it is precisely these terms that were found in \cite{Tyukov:2017uig,Bena:2018mpb} to lead to the strongest tidal forces.  We will see the same phenomenon here.

\section{Tidal forces in superstrata}
\label{Sect:SStides}

The starting point for constructing superstrata is the six-dimensional $(1,0)$ supergravity coupled to two anti-self-dual tensor multiplets.  This theory is obtained by compactifying  type IIB string theory on T$^4$ or K3 and retaining all the fields that are invariant under the rotations on the tangent space of the compactification manifold.  In other words, one only keeps fields that have  no components on the compactifcation manifold, or are proportional to the volume form on this manifold.

The BPS solutions of this six-dimensional theory have been extensively discussed in the literature (see, for example,  \cite{Bena:2015bea,Bena:2016ypk,Bena:2017xbt,Heidmann:2019xrd}).   For BPS solutions, the six-dimensional part of the metric can be written as \cite{Gutowski:2003rg}:
\begin{equation}
d s^2_{6} ~=~-\frac{2}{\sqrt{\cP}}\,(d v+\beta)\,\Big[\, d u+\omega + \coeff{1}{2} \,\mathcal{F}\, (d v+\beta)\, \Big ]+\sqrt{\cP}\,d s^2_4\,.
\label{sixmet}
\end{equation}
For superstrata one takes the metric, $ds_4^2$, on the four-dimensional base, $\cB$, to be that of flat $\IR^4$, and it is most convenient to write this in terms of spherical bipolar coordinates:
 \begin{equation}
 d s_4^2 ~=~ \Sigma \, \left(\frac{d r^2}{r^2+a^2}+ d\theta^2\right)+(r^2+a^2)\sin^2\theta\,d\varphi_1^2+r^2 \cos^2\theta\,d\varphi_2^2\,,
 \label{ds4flat}
\end{equation}
where
 \begin{equation}
\Sigma~\equiv~  (r^2+a^2 \cos^2\theta)     \,.
 \label{Sigdefn}
\end{equation}
The coordinates, $u$ and $v$, are the standard null coordinates, which are related to the canonical time and spatial coordinates via:
\begin{equation}
  u ~=~  \coeff{1}{\sqrt{2}} (t-y)\,, \qquad v ~=~  \coeff{1}{\sqrt{2}}(t+y) \,, \label{tyuv}
\end{equation}
where $y$ is the coordinate around $S^1$ with
\begin{equation}
  y ~\equiv~  y ~+~ 2\pi  R_y \,. \label{yperiod}
\end{equation}

The tensor gauge fields are determined by scalar potentials, $Z_I$, and magnetic two-form fields, $\Theta_I$, on the four dimensional base.  For historical reasons\footnote{These solutions were originally formulated in five dimensions and the fields $Z_3$ and  $\Theta_3$ have become part of the Kaluza-Klein geometry: $Z_3$ has been absorbed in  $\cF$ in (\ref{sixmet}) and $\beta$ is the potential for $\Theta_3$.} the index $I$ takes the values $1,2,4$.     These fluxes and potentials  as well as the function $\cF$, and the one-forms, $\beta$ and $\omega$, on the base $\cB$, are determined by the BPS equations and by requiring regularity.   The details will not concern us here as we will consider classes of solutions that have been constructed elsewhere.

Supersymmetry also fixes the  warp factor, $\cP$, in the metric in terms of the electrostatic potentials:
\begin{equation}
\cP ~\equiv~   Z_1\,Z_2 -  Z_4^2 \,.
\label{Pform}
\end{equation}
The potentials, $Z_1$, $Z_2$ and $\cF$, encode the electric D1, D5 and momentum (P) charges of the system.

\subsection{A terminated-BTZ geometry: the blackened supertube}
\label{ss:BSTs}

Since superstrata are rather cumbersome to work with,  \cite{Bena:2018mpb} introduced a geometry that gives rise to the same tidal forces, but is much easier to construct and analyze: the blackened supertube. This solution has the same throat as a BTZ black hole, but has a nontrivial structure at its bottom. In order to distinguish these generically singular solutions from the smooth capped-BTZ geometries constructed as superstrata and as bubbling geometries, we will refer to them as {\it terminated-BTZ} geometries.

The potentials of the blackened supertube geometry are  \cite{Bena:2018mpb}
\begin{equation}
 Z_1~=~ \frac{Q_1}{\Sigma} \,, \quad  Z_2~=~ \frac{Q_2}{\Sigma} \,, \quad Z_4 ~=~ 0\,;  \qquad \cF ~=~ -  \frac{2\,Q_P}{\Sigma}   \,.
   \label{BSTcharges}
\end{equation}
The supertube has a KKM dipole charge which comes from the non-trivial fibration vector, $\beta$:
 \begin{equation}
\beta ~=~  \frac{R_y \, a^2}{\sqrt{2}\,\Sigma}\,(\, \sin^2\theta\, d\varphi_1 - \cos^2\theta\,d\varphi_2\,)   \,.
 \label{betadefn}
\end{equation}
The exact BPS solution is given by:
\begin{equation}
 \omega ~=~  \omega_0  ~+~ \sqrt{2}\, a^2\, Q_P \,  R_y \, \frac{ \sin^2 \theta  \, \cos^2 \theta }{\Sigma^3}\, \big[ \, (r^2 +a^2)\, d \varphi_1~-~ r^2  \, d \varphi_2 \,\big] \,.
  \label{BSTomega}
\end{equation}
where
\begin{equation}
\omega_0 ~\equiv~  \frac{a^2 \, R_y \, }{ \sqrt{2}\,\Sigma}\,  (\sin^2 \theta  d \varphi_1 + \cos^2 \theta \,  d \varphi_2 ) \,.
\label{angmom0}
\end{equation}

If one sets $Q_P =0$ this solution becomes the smooth, maximally-spinning supertube whose regularity at the {\it supertube locus}, $r = 0$, $\theta =\frac{\pi}{2}$, also requires:  
\begin{equation}
Q_1Q_2 ~=~R_y^2 \,  a^2  \,.
\label{STreg}
\end{equation}

Adding the momentum charge, $Q_P$, creates a singular source at the  supertube locus, as well as closed time-like curves (CTC's) in the immediate vicinity.  Given that such a solution is necessarily singular, one  no longer needs to impose the  condition (\ref{STreg}).  While this solution is certainly pathological  around $r=0$, $\theta =\frac{\pi}{2}$,  we will discuss below how it is still extremely useful as a tool to study tidal forces.

The parameters, $Q_1$, $Q_5$ and $Q_P$ are the supergravity charges of this solution and they are related to the quantized charges, $N_1$, $N_5$ and $N_P$ via \cite{Bena:2015bea}:
\begin{equation}
Q_1 ~=~  \frac{(2\pi)^4\,N_1\,g_s\,\alpha'^3}{V_4}\,,\qquad Q_5 = N_5\,g_s\,\alpha'\,, \qquad Q_P ~=~    \cN^{-1} \, N_P  \,,
\label{QstoNs}
\end{equation}
where $\cN$ is given by:
\begin{equation}
\cN ~\equiv~ \frac{N_1 \, N_5\, R_y^2}{Q_1 \, Q_5} ~=~\frac{V_4\, R_y^2}{ (2\pi)^4 \,g_s^2 \,\alpha'^4}~=~\frac{V_4\, R_y^2}{(2\pi)^4 \, \ell_{10}^8} ~=~\frac{{\rm Vol} (T^4) \, R_y^2}{ \ell_{10}^8} \,.
\label{cNdefn}
\end{equation}
Here, $\ell_{10}$ is the ten-dimensional Planck length and  $(2 \pi)^7 g_s^2 \alpha'^4  = 16 \pi G_{10} ~\equiv~ (2 \pi)^7 \ell_{10}^8$.    The quantity, ${\rm Vol} (T^4)  \equiv (2\pi)^{-4} \, V_4$, is sometimes introduced \cite{Peet:2000hn} as a ``normalized volume'' that is equal to $1$ when the radii of the circles in the $T^4$ are equal to one in Planck units.

The quantized angular momenta can be read-off from the large-$r$ behavior of $\omega$, and are given by
\begin{equation}
j_L ~=~j_R ~=~ \coeff{1}{2} \, \cN \, a^2\,.
\label{angmom}
\end{equation}
%

\subsection{Terminated-BTZ geometries and superstrata}
\label{ss:CappedBTZ}

If one sets $a=0$ in the metric determined by (\ref{BSTcharges}), (\ref{betadefn}) and (\ref{BSTomega}), one obtains the  extremal BTZ metric times that of $S^3$:
\begin{equation}
\begin{aligned}
d s^2_{6} ~=~ \sqrt{Q_1 Q_5}\, \frac{d r^2}{r^2} ~+~ \frac{1}{\sqrt{Q_1 Q_5}}\,\big( -r^2 dt^2 & + r^2 \, dy^2 +  Q_P  \, (dy +  dt)^2 \big)    \\ &~+~  \sqrt{Q_1 Q_5}\, \big( d\theta^2 + \sin^2\theta\,d\varphi_1^2+   \cos^2\theta\,d\varphi_2^2 \big)\,.
\end{aligned}
\label{BTZmet}
\end{equation}
At large $r$, the BTZ metric becomes that of a Poincar\'e AdS$_3$.  For $r<Q_P$, the radius of the $y$-circle stabilizes and the metric looks like a that of a Poincar\'e AdS$_2$ $\times S^1$.  The BTZ horizon is located at $r=0$.  Despite appearances, the BTZ metric is actually that of AdS$_3$ quotiented by a discrete group, and so has constant curvature.

This means that, for small $a$, the blackened supertube metric described in Section \ref{ss:BSTs}, behaves exactly like the BTZ metric  (times  an $S^3$) for $r \gg a$.  In particular, for $r \gg Q_P$, the metric is that of  Poincar\'e AdS$_3$.  If there is a range in which one has $ a^2 \ll r^2 \ll Q_P$ then the geometry will have a long AdS$_2$ $\times S^1$ throat.    Henceforth we will assume that
\begin{equation}
Q_P  ~\gg~a^2\,,
\label{longthroat}
\end{equation}
so that the geometry does indeed have a long BTZ-like throat.

As one approaches $r \sim a$, the geometry ``terminates'' and there is a finite redshift between any point in a smooth region at the cap and any point in the large-$r$, AdS$_3$ region.  For obvious reasons we refer to geometries like this as ``terminated-BTZ'' geometries.  The  unfortunate aspect of the blackened supertube is that it is singular and has closed time-like curves as one approaches $r=0$, $\theta = \frac{\pi}{2}$.

The difference between generic terminated-BTZ geometries and superstrata is that in superstrata the BTZ geometry terminates with a smooth horizonless cap. This is why we generically refer to them as capped-BTZ geometries.  For $r \gg a$, they have exactly the same features as the blackened supertube.  However, in superstrata, the momentum charge is created by a momentum wave traveling along the supertube. There are vast numbers of ways to create such a wave in the CFT  and produce a smooth, horizonless gravity dual \cite{Bena:2015bea,Bena:2016ypk,Bena:2017xbt,Heidmann:2019xrd}.  Such solutions are technically very complicated, but, as was evident from the work of \cite{Bena:2018mpb}, if one wishes to study the tidal effects in the BTZ throat of superstrata, it suffices to work with the much simpler, blackened supertube metric.  We will therefore use this simpler metric and summarize the results found in  \cite{Bena:2018mpb}.

\subsection{Tidal forces in terminated-BTZ geometries}
\label{ss:CappedTides}

The blackened supertube metric has four isometries and they guarantee the following conserved momenta\footnote{As usual with geodesics, these quantities are ``momenta per unit rest mass,'' and so their dimensions must be adjusted accordingly.}:
\begin{equation}
L_1 ~=~ {K_{(1) \mu }} \frac{dx^\mu}{d \tau} \,,  \qquad L_2 ~=~ {K_{(2) \mu }} \frac{dx^\mu}{d \tau} \,,  \qquad    P ~=~ {K_{(3)   \mu }}  \frac{dx^\mu}{d \tau} \,, \qquad E ~=~ - {K_{(4)  \mu }}  \frac{dx^\mu}{d \tau}   \,,
  \label{ConsMom}
\end{equation}
where $K_{(I)}$  are the Killing vectors: $K_{(J)}  = \frac{\partial}{\partial \varphi_J}$, $K_{(3)}  = \frac{\partial}{\partial v }$ and $K_{(4)}  = \frac{\partial}{\partial u}$.  Note that we have reversed the sign of $E$ relative to  \cite{Bena:2018mpb}.

There is also the conserved quantity (\ref{metcons}).  However, this is not enough to determine the geodesic motion.  Instead we use discrete symmetries and fix on the simpler geodesics.  In particular we note that the metric is invariant under $\theta \to  - \theta$ and $\theta \to \pi - \theta$, which means that it is consistent with the geodesic equations to set $\theta =0$ or $\theta = \frac{\pi}{2}$.  Following \cite{Bena:2018mpb} we choose the latter:
\begin{equation}
\theta = \frac{\pi}{2} \,, \qquad \frac{d\theta}{d \tau} ~=~ 0  \,.
  \label{thetamotion}
\end{equation}
For the geodesic to be able to fall from large values of $r$ down to $r=0$, one must take:
\begin{equation}
L_1 = 0\,, \qquad L_2 = 0 \,, \qquad P ~=~ - E  \,.
  \label{CentBarr1}
\end{equation}
For $r \to \infty$, one has
\begin{equation}
\frac{d u}{d\tau}  ~=~ \frac{d v}{d\tau}  ~=~\frac{E\sqrt{Q_{1}Q_{5}}}{r^{2}}  \qquad \Rightarrow \qquad \frac{d t}{d\tau}  ~=~ \frac{E\sqrt{2\,Q_{1}Q_{5}}}{r^{2}}  \,, \quad\frac{d y}{d\tau}  ~=~ 0  \,.
  \label{velinf1}
\end{equation}
Thus the particle has no $y$-velocity at infinity and, for standard time-orientations ($\frac{d t}{d\tau}>0$), one must have
\begin{equation}
E  ~>~   0  \,.
  \label{Epos1}
\end{equation}
Using (\ref{metcons}), the radial motion is determined by
\begin{equation}
\Big(\frac{dr}{d \tau}\Big)^2  ~= \frac{2\,E^2}{  r^2  }\,  \bigg[ \,  (r^2+ a^2) \,\Big(1+\frac{Q_P}{r^2}\Big)- \frac{a^4\, R_y^2}{Q_1 Q_5} \, \bigg]  ~-~  \frac{(r^2+ a^2)}{\sqrt{Q_1 Q_5}} ~ \,.
  \label{Radvel1}
\end{equation}
If the particle is released from rest at $r = r_*$, and if one assume that  $r_*^2 \gg a^2$ and $ \sqrt{Q_1 Q_5} \gg a^2$,  one finds:
\begin{equation}
E^2 ~=~ \frac{r_*^4}{2\, (r_*^2 + Q_P)\,\sqrt{Q_1 Q_5}}  \,.
  \label{Evalue3}
\end{equation}
The magnitude of the generic $E$ for $r_* \gg  \sqrt{Q_P}$ is then:
\begin{equation}
|E|   ~\sim~  \frac{ r_*}{\sqrt{2} \,(Q_1 Q_5)^{1/4}}  \,.
  \label{Evalue4}
\end{equation}
Computing $|\cA|^2$ leads to a  complicated quadratic in $E^2$.  If one sets $a =0$ one arrives at the standard BTZ tidal force result
\begin{equation}
|\cA|_{\rm BTZ}  ~=~ \frac{ \sqrt{2}  }{\sqrt{Q_1 Q_5} }  ~=~ \frac{\sqrt{2}}{\sqrt{N_1 N_5}} \,\frac{\sqrt{{\rm Vol} (T^4)}}{\ell_{10}^4} \,,
  \label{cAthroatBTZ}
\end{equation}
which is always extremely small for large $N_1 N_5$.

Of considerably more interest are terms that dominate when the particle is in the AdS$_2$ throat:
\begin{equation}
a^2 ~\ll~  r^2~\ll~ Q_P \,,  \sqrt{Q_1 Q_5}    \,.
  \label{throat}
\end{equation}
Indeed, the simplest and most effective approach to obtain these terms is to first expand $|\cA|$ for small $a$, and then expand that result for large $Q_1 Q_5$.  One finds the leading term:
\begin{equation}
|\cA|_{\rm throat}  ~\sim~ \frac{ 4\, \sqrt{6} \, a^2 \, Q_P \,E^2 }{r^{6}}  \,\bigg( \, 1  ~+~  \frac{32\, Q_P \, R_y^2 }{3\, Q_1 Q_5}\, \bigg)^{1/2}    \,.
\label{cAcappedBTZ}
\end{equation}
If one takes $r_* \gtrsim  \sqrt{Q_P}$ and implicitly defines $\alpha$ such that $r \equiv a^{(1-\alpha)} Q_P^{\frac{1}{2}\alpha}$ the coefficient in equation (\ref{cAcappedBTZ}) becomes:
\begin{equation}
|\cA|_{\rm throat}  ~\sim~ \frac{ 2\, \sqrt{6} }{\sqrt{Q_1 Q_5}}  \, \bigg(\frac{Q_P}{a^2}\bigg)^{2 - 3 \alpha} 
~=~   \frac{ 2\, \sqrt{6} }{\sqrt{N_1 N_5}}\, \bigg(\frac{N_P}{2\, j_L} \bigg)^{2 - 3 \alpha}  \,\frac{\sqrt{{\rm Vol} (T^4)}}{\ell_{10}^4} \,.  \label{cAcappedBTZval}
\end{equation}
This is the dominant tidal force for $0 < \alpha <\frac{2}{3}$. (For $\alpha > \frac{2}{3}$ the tidal force is dominated by the BTZ result  (\ref{cAthroatBTZ}).)

If one considers superstrata in the Cardy regime
\begin{equation}
 \frac{Q_1 Q_5}{R_y^2} ~\sim~   Q_P  \quad \Leftrightarrow \quad    N_P ~\sim~ N_1 N_5 \,,
  \label{QPorder}
\end{equation}
then the tidal force becomes
\begin{equation}
\begin{aligned}
|\cA|_{\rm throat}  ~\sim~  \frac{1}{a \, R_y}\, \bigg(\frac{Q_P}{a^2}\bigg)^{3\big(\frac{1}{2} -  \alpha\big)} &~=~ \frac{1}{\sqrt{j_L}}\,   \, \bigg(\frac{Q_P}{a^2}\bigg)^{3\big(\frac{1}{2} -  \alpha\big)}  \,\frac{\sqrt{{\rm Vol} (T^4)}}{\ell_{10}^4}\\
& ~=~ \frac{1}{\sqrt{j_L}}\,   \bigg(\frac{N_P}{2\, j_L} \bigg)^{3\big(\frac{1}{2} -  \alpha\big)}  \,\frac{\sqrt{{\rm Vol} (T^4)}}{\ell_{10}^4} \,.
\end{aligned}
\label{tidalorder}
\end{equation}

The microstate geometries with the longest capped BTZ throats have $j_L \sim1$.  One also sees that the tidal forces become large when compared to the compactification scale for $\alpha < \frac{1}{2}$, which corresponds to:
\begin{equation}
r  ~\lesssim~    a^{\frac{1}{2}} \, Q_P^{\frac{1}{4}} \,.
\label{rlimit}
\end{equation}

This was the surprise in \cite{Tyukov:2017uig, Bena:2018mpb}: the tidal force become large, compared to the compactification/Planck scales, at a large distance away from the cap. If one is to measure the length of the  AdS$_2$ throat by naively integrating $\sqrt{g_{rr}}$ from  $r \sim a$ to $r \sim \sqrt{Q_P}$ to , one can see that the location where tides become large is exactly ``half-way down''  the throat.
The goal now is to see how this scrambling might be softened in more generic microstate geometries.

\section{Multi-centered microstate geometries}
\label{sec:multi-centered}

Here we will consider the tidal force on an infalling probe in microstate geometries with a Gibbons-Hawking base. While the previous solutions had non-vanishing angular momentum,   multi-centered solutions with a large number of centers can be arranged to have small, or even vanishing, multipole moments. We would like to understand how this affects the tidal force.

Because of their complexity, it is not possible to work directly with solutions made of a large number of GH points. We focus in this section on a simple model with five GH points and a vanishing $SU(2)_L$ angular momentum, $J_L$, similar to the ``pincer movement''  described in \cite{Bena:2006kb}. In the scaling limit, it can be seen as a microstate geometry corresponding to two concentric black rings \cite{Gauntlett:2004qy}.

\subsection{A smooth solution with 5 GH centers}
\label{ss:a_smooth_solution_with_5_gh_centers}

The solution we consider is most easily written in terms of the five-dimensional GH formulation, except that we are going to uplift this to six-dimensions and write everything in terms of six-dimensional quantities so as to facilitate comparisons of tidal forces.  We follow the standard procedure as outlined in \cite{Bena:2006kb, Bena:2007kg, Warner:2019jll}.

The metric is still (\ref{sixmet}), but the four-dimensional part, $ds_4^2$, is the Gibbons-Hawking metric written in cylindrical polar coordinates:
\begin{equation}
ds_4^2 ~=~ V^{-1} \, \big( d\psi + A \big)^2  ~+~ V\, (d\rho^2 + \rho^2 d\phi^2 + dz^2) \,.
\label{GHmetric}
\end{equation}

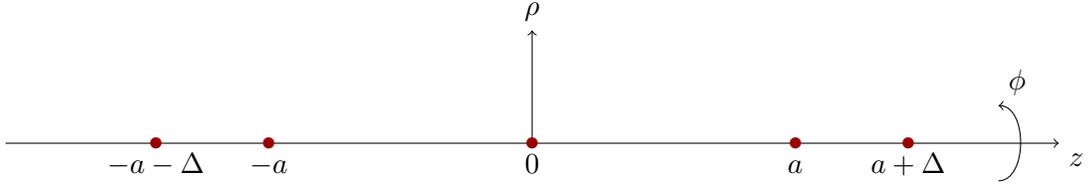
\begin{figure}
\centering
\begin{tikzpicture}[scale=1]
        \draw[->] (-7,0) -- (7,0);
        \draw (7,0) node[below right] {$z$};
        \draw[->] (0,0) -- (0,1.5);
        \draw (0,1.5) node[above] {$\rho$};
        \draw[->] (6.2,-.5) to[bend right=90] (6.2,.5);
        \draw (6.2,.5) node[above right] {$\phi$};

        \draw (0,0) node {\color{darkred}$\bullet$};
        \draw (0,0) node[below] {$0$\vphantom{$\Delta$}};

        \draw (3.5,0) node {\color{darkred}$\bullet$};
        \draw (3.5,0) node[below] {$a$\vphantom{$\Delta$}};

        \draw (-3.5,0) node {\color{darkred}$\bullet$};
        \draw (-3.5,0) node[below] {$-a$\vphantom{$\Delta$}};

        \draw (5,0) node {\color{darkred}$\bullet$};
        \draw (5,0) node[below] {$a+\Delta$};

        \draw (-5,0) node {\color{darkred}$\bullet$};
        \draw (-5,0) node[below] {$-a-\Delta$};
\end{tikzpicture}
\caption{The model we consider has  5 Gibbons-Hawking points arranged in a $\ZZ_2$ symmetric manner along the z-axis.  The locations of the points are determined by  two scale parameters, $a$ and $\Delta$.}
\label{fig:ghpoints}
\end{figure}

To construct a solution with vanishing $SU(2)_L$ angular momentum, we choose five Gibbons-Hawking points aligned on the z-axis, and symmetric under the $\ZZ_2$ transformation $z \to -z$. Specifically, their distances to the origin are respectively denoted by $z_1 = -\Delta - a$, $z_2 = -a$, $z_3 = 0$, $z_4 = a$ and $z_5 = a + \Delta$ (see Fig. \ref{fig:ghpoints}).

We also choose the Gibbons-Hawking charges to be $\ZZ_2$ symmetric :
\begin{equation}
    V ~=~ \frac 1{\hat r} + q \qty( \frac 1{\hat r_1} - \frac 1{\hat r_2} - \frac 1{\hat r_4} + \frac 1{\hat r_5} )
    \label{v5points}
\end{equation}
where $\hat r_j~\equiv~ \sqrt{\rho^2 + (z - z_j)^2}$ denotes the  distance to the $j$-th point in the $\IR^3$ base of the GH space. The potential, $A$, is then fixed by requiring $\vec\nabla\times \vec A = \vec \nabla V$ and $A \to 0$ along the positive $z$-axis.

We now introduce the harmonic functions
\begin{align}
    K^I &~=~ k^I \qty(\frac 1{\hat r_2} + \frac 1{\hat r_4}) + \tilde k^I \qty(\frac 1{\hat r_1} + \frac 1{\hat r_5})
    \\
    L_I &~=~ \delta_I^{\,3} -\frac {\abs{\epsilon_{IJK}}}{2 q} \qty(-  k^J k^K \qty(\frac 1{\hat r_2} + \frac 1{\hat r_4}) +  \tilde k^J \tilde k^K \qty(\frac 1{\hat r_1} + \frac 1{\hat r_5}))
    \\
    M &~=~ -(k^3 + \tilde k^3) + \frac 1{2q^2} \qty(k^1 k^2 k^3 \qty(\frac 1{\hat r_2} + \frac 1{\hat r_4}) + \tilde k^1 \tilde k^2 \tilde k^3 \qty(\frac 1{\hat r_1} + \frac 1{\hat r_5}))
\end{align}

for $I = 1, 2, 3$, and define the warp factors appearing in the metric
\begin{equation}
  \stP ~=~ Z_1 Z_2 \qq{and} \mathcal{F} ~=~ 2(1- Z_3)\,, \qquad Z_I ~\equiv~ L_I + \frac 12 \abs{\epsilon_{IJK}} V^{-1} K^J K^K \,. \label{Zdef}
\end{equation}

The remaining variables take the form
\begin{align}
  \beta &~=~ V^{-1} K^3 (\dd{\psi} + A) + \sigma  \label{eq:betaomega}\\
  \omega &~=~ \mu  (\dd{\psi} +A) + \varpi \label{eq:betaomega2}
\end{align}
with
\begin{equation}
  \mu ~=~ V^{-2} K^1 K^2 K^3 + \frac 12 V^{-1} K^I L_I + M \label{mudef}
\end{equation}
and
\begin{align}
  \vec \nabla \times \sigma &~=~ -\vec \nabla K_3 \\
  \vec \nabla \times \varpi &~=~ V \vec \nabla M - M \vec \nabla V + \frac12 (K^I \vec \nabla L_I - L_I \vec \nabla K^I)
\label{curls}
\end{align}

These last equations have standard solutions (see, for example, \cite{Bena:2005va,Berglund:2005vb,Bena:2007kg,Warner:2019jll}). The constant $d\phi$ components in $\sigma$ and $\varpi$ are fixed by requiring that $\beta$ and $\omega$ vanish as $\hat r \to \infty$.  We will also add a constant multiple of  $d\psi$ to  $\beta$ so as to cancel the constant part of $K_3/V$ in (\ref{eq:betaomega}) as $\hat r \to \infty$.

The coefficients in the harmonic forms have been chosen to ensure that the resulting metric is everywhere smooth (it has no Dirac strings) providing it verifies the ``bubble equations'' \cite{Bena:2005va,Berglund:2005vb,Bates:2003vx}. In  our geometries these equations are:
\begin{align}
    \begin{split}
         a \Delta  (2 a+\Delta ) q^2 \big(k^3 &+ 2 q (k^3 + \tilde k^3)\big) ~=~ k^1 k^2 k^3 \left(2 a^2 q+ 2 a \Delta  (q+1)+\Delta^2\right)\\
         & + 2 a q (a+\Delta) \left(\tilde{k}^1 \left(\tilde{k}^2+k^2\right) \left(\tilde{k}^3+k^3\right)+k^1 \tilde{k}^2 \left(\tilde{k}^3+k^3\right)+k^1 k^2 \tilde{k}^3\right)
    \end{split} \\
    \big(k^1 k^2 k^3 - a q^2 (k^3 &+ \tilde k^3)\big) (a+\Delta) ~=~ - \tilde{k}^1 \tilde{k}^2 \tilde{k}^3 a
\label{bubbleeqns}
\end{align}
One can use the second equation to express $\Delta$ as a function of $a$, and then replace it in the first equation to obtain $a$ as the root of a third-order polynomial.

\subsection{Asymptotic expansion of the solution}
\label{ss:asymptotic_expransion}

Although this model is much simpler than a generic multi-center solution, it still has a level of complexity that prohibits the exact computation of $|\cA|$.
We therefore use an approximation of the solution, and make sure it is good enough to contain the terms we are looking for. We are mostly interested in the behavior of the tidal force in the AdS$_2$ throat, for which
\begin{equation}
  a, \Delta ~\ll~ \hat r ~<~ Q, Q_P \,.
\end{equation}

An expansion of the metric in the radial distance would not be trivial here because of the constraint that $r$ must be smaller than the charges. A simpler choice is to expand the metric in small $a$ and small $\Delta$. We will then write $\Delta \equiv \delta \, a$ with $\delta < 1$, and expand the metric in $a$.

The form of the expansion in spherical coordinates is as follows ($V$ and $A$ are given by Legendre polynomials) :

\begin{align}
\mathcal{F} ={}& -\frac{Q_P}{4 \hat r}-\frac{a^2 Q_P (3 \cos^2 \theta-1) f_{3}}{\hat r^3}-\frac{a^4 Q_P}{\hat r^5}\left(\sum_{k=0}^2 f_{5, 2k} \cos (2k\theta)\right) + O\qty(a^5) \nonumber
\\
 \sqrt{\cP} ={}& \frac{Q}{4 \hat r}+\frac{a^2 Q \left(3 \cos ^2\theta-1\right) p_{3}}{\hat r^3}+\frac{a^4 Q }{\hat r^5} \left(\sum_{k=0}^2 p_{5,2k} \cos(2k \theta)\right) + O\qty(a^5) \nonumber
\\
V = {}& \frac{1}{\hat r}+\frac{a^2 \left(3 \cos ^2\theta-1\right) v_{3}}{2 \hat r^3}+\frac{a^4 \left(35 \cos ^4\theta-30 \cos ^2\theta+3\right) v_{5}}{8 \hat r^5} + O\qty(a^5) \nonumber
\\
A_\phi ={}& (\cos\theta-1)-\frac{3 a^2 \cos \theta \sin ^2\theta \, v_{3}}{2 \hat r^2}+\frac{a^4 \sin \theta \left(60 \cos \theta \sin \theta-140 \cos ^3\theta  \sin \theta \right) v_{5}}{32 \hat r^4} + O\qty(a^5) \nonumber
\end{align}
\begin{align}
\beta_\psi ={}& \frac{a^2 B \left(3 \cos ^2 \theta -1\right)}{\hat r^2}+\frac{a^4 B}{\hat r^4}\left(\sum_{k=0}^2 b_{4,2k} \cos(2k \theta)\right) + O\qty(a^5) \nonumber
\\
\beta_\phi ={}& \frac{a^2 B (4 \cos\theta -3 \cos (2 \theta )-1)}{2 \hat r^2} + \frac{a^4 B (\cos\theta-1)}{\hat r^4} \left(\sum_{k=0}^3 \tilde{b}_{4,k} \cos(k \theta)\right) + O\qty(a^5)
\label{expansion}\\
\omega_\psi ={}& \frac J{\hat r} + \frac{a^2 J \qty(-1+3\cos^2\theta) k^\psi_3}{\hat r^3} + \frac{a^4 J}{\hat r^5} \qty(\sum_{k=0}^4 k^\psi_{5, 2k} \cos (2k\theta)) + O\qty(a^5) \nonumber
\\
\begin{split}
    \omega_\phi ={}& \frac {J(-1+\cos \theta)}{\hat r} - \frac{3 a J \cos\theta \sin^2\theta k^\phi_2}{\hat r^2} + \frac{a^2 J \qty(\qty(-1+\cos\theta)\qty(-1+3\cos^2\theta) k^\psi_3 - 3 \cos\theta \sin^2\theta k^\phi_3)}{\hat r^3}
    \\
    & - \frac{5 a^3 J \qty(9\cos\theta + 7 \cos(3\theta))\sin^2\theta k^\phi_4 }{8 \hat r^4}
    \\
    & + \frac{a^4 J}{8 \hat r^5}\Biggl(8(-1+\cos\theta)\biggl(\sum_{k=0}^4 k^\psi_{5, 2k} \cos (2k\theta)\biggr) - 6\cos\theta (1+3\cos(2\theta)) \sin^2\theta k^\psi_3 v_3 \\
    & - 5(9\cos\theta + 7\cos(3\theta))\sin^2\theta k^\phi_5\Biggr) + O\qty(a^5)
\end{split} \nonumber
\end{align}

This expansion depends on the asymptotic charges $Q = \sqrt{Q_1 Q_5}$, $Q_P$, $J \equiv J_R$ and $B$, as well as on several dimensionless parameters, $f_k$, $p_k$, $v_k \dots$.

By expanding the exact solution, defined by (\ref{GHmetric})--(\ref{curls}), in powers of $a$, these parameters can be matched onto expressions involving $q$, the charges $k^I$ and $\tilde{k}^I$, and the distance ratio $\delta$.

The ``unusual'' charge, $B$, is expressed in terms of these parameters as
\begin{equation}
    B ~=~ \big( 1 - 2\, q\, \delta (2 + \delta)\big)\,  \qty( k^3 + \tilde k^3 ) ~+~ \delta (2 + \delta)\, \tilde k^3 \,.
\end{equation}

Note that we need to expand up to order $a^4$, because the dominant term in the tidal force in superstrata and in  terminated-BTZ geometries was of this order.

\subsection{Tidal forces} 
\label{ss:tidal_forces}

The computation of the tidal forces of an infalling particle in this geometry is performed in much the same way as in Section \ref{ss:CappedTides}. We restrict ourselves to geodesics along the z-axis as they are simpler to compute, since they have no velocity along the $\phi$ and $\rho$ directions.

The Killing vectors of the metric are
\begin{equation}
\cK^{(1)}  = \frac{\partial}{\partial u}  \,, \qquad \cK^{(2)}  = \frac{\partial}{\partial v}  \,, \qquad \cK^{(3)}  = \frac{\partial}{\partial \psi}  \,, \qquad \cK^{(4)}  = \frac{\partial}{\partial \phi}  \,.
  \label{Kvecs}
\end{equation}

They are associated to the following conserved quantities :
\begin{equation}
E ~=~ - {\cK^{(1)}}_ \mu \frac{dx^\mu}{d \tau} \,,  \qquad P_3 ~=~ {\cK^{(2)}}_ \mu \frac{dx^\mu}{d \tau} \,,  \qquad    P_1 ~=~{\cK^{(3)}}_ \mu \frac{dx^\mu}{d \tau} \,, \qquad P_2 ~=~ {\cK^{(4)}}_ \mu \frac{dx^\mu}{d \tau}   \,.
  \label{ConsMom2}
\end{equation}

We are looking for geodesics that fall down to $\hat r=0$. To remove the centrifugal barriers, we need to impose
\begin{equation}
  P_1 = P_2 = 0 \qq{and} E = P_3 \,.
  \label{geod_condition4}
\end{equation}
This fixes all components of the velocity but one:
\begin{align}
    \dv{u}{\tau} ~&=~ \frac E{\sqrt{\cP}} \, \qty(V \mu \qty(\mu + \frac{K^3}V) \,-\, \cP (1-\mathcal{F}))
    \\
    \dv{v}{\tau} ~&=~ -E \sqrt{\cP} \,+\, \frac E{\sqrt{\cP}}\, K^3 \qty(\mu + \frac{K^3}V)
    \\
    \dv{\psi}{\tau} ~&=~ - \frac E{\sqrt{\cP}} \, V\, \qty(\mu + \frac{K^3}V)
\end{align}
The radial velocity is then determined using the metric condition (\ref{metcons}), which translates to
\begin{equation}
    \sqrt{\cP}\, V \, \qty(\dv{z}{\tau})^2 \, +\, \frac{E^2}{\sqrt{\cP}} \qty( V \qty(\mu +\frac{K^3}V)^2 - \cP\, (2-\mathcal{F}) ) ~=~ -1 \,.
\end{equation}

Note that if one lets the geodesic start from a large distance $\hat r_*$, by requiring $\frac{\dd{r}}{\dd{\tau}}|_{\hat r = \hat r_*} ~=~ 0$ one finds that its energy scales like
\begin{equation}
  \abs{E} ~\sim~ 2 \sqrt{\frac {\hat r_*}{Q}} \,,
  \label{Epart4}
\end{equation}
which is similar to the previous result (\ref{Evalue4}) (the distance $\hat r_*$ is here in the coordinates of the $\IR^3$ base of the Gibbons-Hawking space).

It is then possible to compute the tidal force through the formula (\ref{cAnorm}). We look for the dominant contribution well inside the AdS$_2$ throat, in the regime
\begin{equation}
     a ~\ll~ z ~\ll~ Q, Q_P \,.
\label{regime_halfway}
 \end{equation}

The leading term in $E^2$ of the norm of the tidal tensor is determined to be
\begin{equation}
    \abs{\mathcal A} ~\sim~ \frac{C a^2 Q_P E^2}{\hat{r}^4}
    \label{leadingAterm4}
\end{equation}
where $C$ is a constant depending on the parameters of the expansion (\ref{expansion}), and on $\delta$. It contains a great number of terms, but if we focus on the terms independant of $J$, we find
\begin{equation}
    C^2 \supset 18\, \frac{B^2 Q_P}{Q^2} \, +\, \frac{27}{32}\, v_3^2 \,-\, \frac 92 \, v_3\, (2 p_3 + f_3) \,+\, 72\, p_3\, (p_3 - f_3) \,+\, 54\, f_3^2 \,.
\end{equation}

There are other terms with lower powers of $\hat r ^{-1}$ in the tidal force, but they are sub-leading because they also have lower powers of $Q$ or $Q_P$, especially when one takes into account the fact that $E$ is large.

This result is to be compared to the one for superstrata and for the terminated-BTZ geometry (\ref{cAcappedBTZ}). We recall that $a$ and $\hat r$ are distances in GH coordinates, they are related to the spherical bipolar coodinates through $a \approx \frac 14 a^2_{\rm{s.b.}}$ and $\hat r \approx \frac 14 r^2_{\rm{s.b.}}$. The tidal force we find here is suppressed by a factor $a/\hat r$ relative to the result in  superstrata and in the terminated-BTZ geometry.

Using  (\ref{Epart4}) with $\hat r_* = Q_P$ and setting  $4 \hat r = (4 a)^{(1-\alpha)} Q_P^{\alpha}$, one obtains
\begin{equation}
\abs{\cA} ~\approx~ \frac{C}{Q}  \,\qty(\frac{Q_P}{4 a})^{2(1-2\alpha)} ~\approx~  \frac{ C }{\sqrt{N_1 N_5}}\, \bigg(\frac{N_P}{2\,\hat j_L} \bigg)^{2(1-2\alpha)}  \,\frac{\sqrt{{\rm Vol} (T^4)}}{\ell_{10}^4} \,.
\label{E4leadingf3}
\end{equation}
where we have introduced $\hat j_L = 2 \mathcal{N} a$ as a parameter that measures the angular momentum of each bubble. This expression should be compared with (\ref{cAcappedBTZval}).  Here the cross-over between the tidal forces from the constant-curvature background to the dominance of multipoles comes at $\alpha =\frac{1}{2}$, as opposed to $\alpha =\frac{1}{3}$.  Moreover, if one also has $Q_p$ of the order given in  (\ref{QPorder}), one finds
\begin{equation}
\abs{\cA} ~\approx~ C \sqrt{\frac{ 1 }{\hat j_L}}\, \bigg(\frac{N_P}{2\,\hat j_L} \bigg)^{\frac{1}{2}(3-8\alpha)}  \,\frac{\sqrt{{\rm Vol} (T^4)}}{\ell_{10}^4} \,.
\label{E4leading4}
\end{equation}
If the angular momentum, $\hat j_L$, on the individual black rings is small, then the tidal forces now become large for $\alpha  < \frac{3}{8}$.

Cancelling the full $SU(2)_L$ angular momentum, $J_L$, thus softens the tidal forces, but these forces still become large a long distance away from the cap.
 
\subsection{A slight asymmetry} 
\label{ss:a_slight_asymmetry}

Given the previous result, we would like to know if it a more generic configuration will reproduce  the large tidal force of the superstrata and the terminated-BTZ geometries discussed in Section \ref{Sect:SStides}. We therefore  introduce a small asymmetry in the distribution of GH points.

The simplest way to add an asymmetry to the five-center solution is to change one of the magnetic charges of a GH point. Our choice is to modify the $K^3$ charge of the fifth GH point: $\tilde k^3 \to \tilde k^3 (1+\epsilon)$ (the charge of the first GH point stays $\tilde k^3$).

This change in the fluxes modifies the bubble equations, which in term give rise to a small change of the distances between centers. However, in order to see how the asymmetry affects the tidal force one can still use the asymptotic expansion of the solution where the distances between the centers have not changed. This solution will have regions with closed timelike curves in the vicinity of the GH centers but, using the philosophy of  \cite{Bena:2018mpb}, one can still use it to understand the generic terms present in the tidal force on an incoming geodesic.

The form of the asymptotic expansion is modified: one has to add terms in odd powers of $a$ in $\sqrt{\cP}$, $\omega_\psi$, and $\beta$, and add a term to $\omega_\phi$. These new terms are all proportional to $\epsilon$. The expansion of $\mathcal{F}$ is unchanged.

The same procedure described in the previous sub-section is then used to compute the tidal force on a geodesic along the z-axis. The Killing vectors and conserved quantities are the same. In the regime (\ref{regime_halfway}) we look for the dominant terms in $\epsilon$ as $\epsilon\to 0$. We do not consider the terms independent of $\epsilon$ since they are identical to the ones found in the previous subsection. The terms proportional to $E^2$ are
\begin{equation}
  \abs{\mathcal{A}} ~\sim~  \sqrt{\epsilon}\,\frac{C_1\, a^{3/2}\, Q_P^{1/2} E^2}{\hat r^3} ~+~ \epsilon \, \frac{C_2\, a\, Q_P E^2}{\hat r^3}
  \label{asymmetry_tidal4}
\end{equation}
where $C_1$ and $C_2$ are once again constants that depend on the parameters appearing in the expansion (\ref{expansion}) and on $\delta$. While the first term dominates as $\epsilon \to 0$, the second is more relevant to our discussion since it has a greater power of $Q_P$.

Looking at the terms independent of $J$ in $C_2$, we find
\begin{equation}
    C_2^{\hspace{0.5ex}2} \supset \frac{B^2 Q_P}{2 Q^2} \,+\, 2 \, p_2^2 \,.
\end{equation}

The second term in the tidal force (\ref{asymmetry_tidal4}) is of the same order as (\ref{cAcappedBTZ}). As it is proportional to $\epsilon$, this confirms the fact that this term comes from the non-vanishing angular momentum $J_L$. This is further confirmed in the next section in which we reproduce the same result using a solution with two concentric black rings.



\section{Blackened bubbles}
\label{sec:BlackenedBubbles}

The goal of this section is to use black rings in much the same manner as one can use blackened supertubes to provide a simpler model of the tidal forces in a scaling, multi-centered bubbled geometry.  Specifically, we are going to replace the pairs of GH points used in Section \ref{sec:multi-centered} by concentric black rings  \cite{Gauntlett:2004qy, Gauntlett:2004wh}.  The idea is that the interesting tidal effects arise at a significant distance away from where one encounters the individual black rings, or pairs of GH points, and so this will provide a sufficiently good approximation to the tidal effects in the throat of a multi-centered microstate geometry.

\subsection{Two concentric black rings}
\label{ss:towBBs}

\def\apoint{c_1}
\def\bpoint{c_2}
\def\rapoint{{\hat r}_1}
\def\rbpoint{{\hat r}_2}

The solution we consider is still written in the Gibbons-Hawking formalism. The four-dimensional base is given by the metric (\ref{GHmetric}). However we will reduce the base to  $\IR^4$  by taking $V$ and $A$ to be:
\begin{equation}
 V  ~=~ \frac{1}{{\hat r}} \,, \qquad  A  ~=~ \frac{z}{{\hat r}} \, d \phi\,
\label{VAform}
\end{equation}

The underlying harmonic functions are:
\begin{equation}
\begin{aligned}
K^1 ~=~  & K^2 ~=~ \frac{k}{\rapoint} +  \frac{k}{\rbpoint}\,, \qquad   K^3~=~ \frac{k_3}{\rapoint} +  \frac{k_3}{\rbpoint} \,, \qquad L_1 ~=~ L_2 ~=~ \frac{Q}{ \rapoint } ~+~  \frac{Q}{ \rbpoint}\,, \\  L_3~=~ & 1~+~  \frac{Q_P}{\rapoint} ~+~  \frac{Q_P}{\rbpoint} \,, \qquad 
M~=~  -k_3 ~+~  \frac{k_3}{2}\, \bigg(\frac{\apoint}{\rapoint} ~+~  \frac{\bpoint}{\rbpoint}  \bigg) \,, 
\end{aligned}
\label{KLMform}
\end{equation}
where
\begin{equation}
 \rapoint  ~\equiv~ \sqrt{\rho^2 + (z- \apoint)^2} \,, \qquad   \rbpoint  ~\equiv~ \sqrt{\rho^2 + (z+\bpoint)^2} \,,
\label{VAform2}
\end{equation}
and $\apoint, \bpoint  >0$. Note that the poles of the $M$ harmonic function have already been chosen such that the configuration solves the bubble equations, and hence has no Dirac-Misner strings.  We also recall the warp factors and angular momentum function along the GH fiber defined in (\ref{Zdef},\ref{mudef}):
\begin{equation}
\begin{aligned}
Z_1 ~=~  & Z_2 ~=~   \frac{K^1 \, K^3}{V} ~+~   L_1  \,,  \qquad  Z_3 ~=~    \frac{K^1 \, K^2}{V} ~+~   L_3  \\
\mu ~=~ & V^{-2} K^1 K^2 K^3 ~+~ \coeff{1}{2}\, V^{-1}\big( K^1 L_1 + K^2 L_2+ K^3 L_3  \big) ~+~ M  \,.
\end{aligned}
\label{Zmuform}
\end{equation}
Furthermore, the angular momentum along the $\IR^3$ base of the GH space, satisfying (\ref{curls}) is:
\begin{equation}
 \varpi  ~=~\frac{k_3}{2\,{\hat r} }   \bigg( \frac{1}{\rapoint} \,\big(\rho^2 +(z-\apoint +\rapoint)(z-{\hat r})\big) ~-~\frac{1}{\rbpoint} \,\big(\rho^2 +(z+\bpoint -\rbpoint)(z+{\hat r})\big)   \bigg)\, d \phi\,.
\label{varpi}
\end{equation}

This solution represents two concentric black rings, one wrapping the $\psi$-fiber at $\rho =0$, $z=\apoint$ and the other wrapping the $\psi$-fiber at $\rho =0$, $z=-\bpoint$. These rings have the same dipole charges $(k, k, k_3)$ and the same Page charges $(Q, Q, Q_P)$ \cite{Chowdhury:2013pqa}. Their asymmetry only comes from the coefficients of the poles in the harmonic function,  $M$, and these constants encode the intrinsic  angular momenta of the rings \cite{Bena:2004tk}.

These ring angular momenta in turn control the locations of the rings on the $z$-axis. Note that the full supergravity angular momentum charges are not the same as the ring angular momenta:
\begin{equation}
J_R ~=~ 8 \,(8 \, k\, Q + 4 \, k_3\, Q_P+ k_3(\apoint+\bpoint) + 16\, k^2 \, k_3) \,, \qquad  J_L ~=~ 8 \,k_3\,(\apoint-\bpoint) \,.
\label{JLJR}
\end{equation}

Finally, we observe that the change of coordinate that takes the origin, ${\hat r}=0$, and the point $\rho =0$, $z=\apoint$ to the spherical bipolars is defined by \cite{Bena:2017geu}:
\begin{equation}
\hat r ~=~\coeff{1}{4}( r^2 + a^2 \cos^2 \theta )\,, \qquad  \rapoint ~=~ \coeff{1}{4}( r^2 + a^2 \sin^2 \theta)  \,,  \qquad \apoint ~=~ \coeff{1}{4}\,a^2  \,.
\label{bipolarchange}
\end{equation}
To find the conserved charges one must expand the warp factors at infinity
\begin{equation}
Z_{1,2} ~\sim \frac{2 (Q+2 k k3)}{\hat r}  ~\sim ~  \frac{8 (Q+2 k k3)}{r^2}  \,, \qquad   Z_3  ~\sim ~  1 + 
\frac{2 (Q_P+2 k^2)}{\hat r}  ~\sim ~ 1+ \frac{8 (Q_P+2 k^2)}{r^2}  \,.
\label{ZFasymp}
\end{equation}

If one dualizes the five-dimensional supergravity solution encoded by these functions to the duality frame in which the charges of the black rings correspond to D1 branes, D5 branes and momentum along the common D1-D5 direction,  there are two ways to obtain a six-dimensional solution.  These were discussed in detail in Appendix B of  \cite{Bena:2016agb}. The first way\footnote{This corresponds to  Reduction 1 in the language of \cite{Bena:2016agb}.} gives rise to the six-dimensional supergravity parameters \cite{Bena:2008wt}:
\begin{equation}
  \cP ~=~Z_1 \, Z_2 \,, \qquad \cF = - Z_3 \,,
\label{PFform}
\end{equation}
\begin{equation}
\beta~=~\frac{K^3}{V} \,  (d \psi +A)   ~-~ k_3\, \bigg( \frac{(z-\apoint)}{\rapoint} + \frac{(z+\bpoint)}{\rbpoint}    \bigg)\, d \phi \,, \qquad \omega ~=~ \mu \,  (d \psi +A)   ~+~ \varpi\,.
\label{betaomform}
\end{equation}
and this leads to
\begin{equation}
u ~=~t \,, \qquad v~=~t+y \,.
\label{identifications1}
\end{equation}
This is different from the uplift used in Section \ref{Sect:SStides}, and which gives the coordinates\footnote{This corresponds to Reduction 2 in the language of \cite{Bena:2016agb}.} (\ref{tyuv}). As explained in  \cite{Bena:2016agb} the two ways of relating five-dimensional and six-dimensional solutions are related by a coordinate transformation in six dimensions. The expression (\ref{betaomform}) for $\beta$, is also in a different gauge to the one given in (\ref{betadefn}).

\subsection{Frames and geodesics}
\label{ss:FramesGeos}

To find convenient set of frames, we first note that one can write (\ref{sixmet}) as
\begin{equation}
d s^2_{6} ~=~ -\frac{1}{Z_3 \,\sqrt{\cP}}\,(d u+\omega)^2 ~+~ \frac{Z_3}{\sqrt{\cP}}\, \big(\,(d v+\beta)  ~-~ Z_3^{-1} (d u+\omega) \,  \big)^2+\sqrt{\cP}\,d s^2_4\,.
\label{sixmet2}
\end{equation}
Since $Z_3 \to 1$ and $\sqrt{\cP} \sim r^{-2}$ as $r \to \infty$,  the coordinate identifications  (\ref{identifications1}) become clear.

We therefore choose frames:
\begin{equation}
e^{0}~\equiv~  Z_3^{-\frac{1}{2}} \,\cP^{-\frac{1}{4}}  \, (d u+\omega)  \,,\qquad e^{1}~\equiv~ \cP^{-\frac{1}{4}}  \, \Big(Z_3^{\frac{1}{2}} \,(d v+\beta)  ~-~ Z_3^{-\frac{1}{2}} \, (d u+\omega) \,  \Big) \,,  \qquad e^{j+1}~\equiv~\cP^{\frac{1}{4}} \, \hat e^j \,,
\label{sixframes}
\end{equation}
for  $ j=1,2,3,4$, where $\hat e^j $ are orthonormal frames on $ds_4^2$.  In Appendix \ref{app:Curvatures} we compute the frame connections and curvature for this set of frames, using a generic four-dimensional base metric (assuming that all quantities are independent of both $u$ and $v$).  These explicit formulae greatly streamline the Mathematica computations later.

Here, however, we use the following four-dimensional frames, based on the metric given by (\ref{GHmetric}) and (\ref{VAform}):
\begin{equation}
\hat e^{1}~\equiv~  V^{-\frac{1}{2}}\, ( d\psi + A)\,,\qquad  \hat e^{2}~\equiv~  V^{\frac{1}{2}} \, d\rho\,,\qquad  \hat e^{3}~\equiv~  V^{\frac{1}{2}} \, \rho\, d \phi \,,\qquad  \hat e^{4}~\equiv~ V^{\frac{1}{2}} \, dz  \,.
\label{fourframes}
\end{equation}

The Killing vectors are now:
\begin{equation}
\cK^{(1)}  = \frac{\partial}{\partial u}  \,, \qquad \cK^{(2)}  = \frac{\partial}{\partial v}  \,, \qquad \cK^{(3)}  = \frac{\partial}{\partial \psi}  \,, \qquad \cK^{(4)}  = \frac{\partial}{\partial \phi}  \,,
  \label{Kvecs2}
\end{equation}
with the associated conserved quatities
\begin{equation}
E ~=~ - {\cK^{(1)}}_ \mu \frac{dx^\mu}{d \tau} \,,  \qquad P_3 ~=~ {\cK^{(2)}}_ \mu \frac{dx^\mu}{d \tau} \,,  \qquad    P_1 ~=~{\cK^{(3)}}_ \mu \frac{dx^\mu}{d \tau} \,, \qquad P_2 ~=~ {\cK^{(4)}}_ \mu \frac{dx^\mu}{d \tau}   \,.
  \label{ConsMom3}
\end{equation}

It is useful to define frame velocities:
\begin{equation}
v^{a}  ~\equiv~ { e^{a}}_\mu \,  \frac{dx^\mu}{d \tau}  \,, \qquad a=0,1\ldots,5 \,.
  \label{framvels1}
\end{equation}
One can then use the conserved quantities to write:
\begin{equation}
\begin{aligned}
v^{0}  & ~=~ Z_3^{-\frac{1}{2}} \,\cP^{\frac{1}{4}} \, (Z_3 E- P_3)  \,, \qquad v^{1}   ~=~ Z_3^{-\frac{1}{2}} \,\cP^{\frac{1}{4}} \, P_3  \,, \qquad  v^{2}   ~=~ \cP^{-\frac{1}{4}} \, V^{\frac{1}{2}} \, \bigg( E \mu + P_1 - \frac{K^3}{V} P_3 \bigg)\\
v^{3}  & ~=~\cP^{\frac{1}{4}} \, V^{\frac{1}{2}} \, \frac{d\rho }{d \tau}   \,, \qquad  v^{4}    ~=~\cP^{\frac{1}{4}} \, V^{\frac{1}{2}} \,\rho\, \frac{d\phi }{d \tau}   \,, \qquad  v^{5}   ~=~\cP^{\frac{1}{4}} \, V^{\frac{1}{2}} \, \frac{dz }{d \tau}     \,.
\end{aligned}
  \label{framvels}
\end{equation}

We are going to consider geodesics that have no angular motion in $y$, $\psi$ and $\phi$ at infinity.  This means
\begin{equation}
P_1~=~ P_2~=~ P_3~=~ 0   \,.
  \label{noangmom}
\end{equation}
This differs from  (\ref{CentBarr1}) and (\ref{geod_condition4}) because we are using six-dimensional coordinates 
\eqref{identifications1}, and not (\ref{tyuv}).

 We are also going to consider the simplest possible infalling geodesics:  those that fall along the $z$-axis from $z \gg \apoint$.  It is easy to check that this is consistent with the geodesic equations.  As a result we also have
\begin{equation}
\frac{d\rho }{d \tau} ~=~ \frac{d\phi }{d \tau}~=~  0   \,,
  \label{noangmom2}
\end{equation}
 along the entire geodesic.  With these choices (\ref{metcons}) becomes
\begin{equation}
-(v^{0})^2~+~ (v^{2})^2 ~+~ (v^{5})^2 ~=~\sqrt{\cP} \, V \, \bigg(\frac{dz }{d \tau}\bigg)^2 ~+~ \frac{ E^2}{\sqrt{\cP} } \, \big( V\, \mu^2~-~ Z_3 \,\cP \big) ~=~ -1  \,,
  \label{metcons2}
\end{equation}
which determines the proper $z$ velocity.   One can use this to verify that if the probe is released from rest at $z \sim Q_P$, then one has
\begin{equation}
E  ~\sim~ \sqrt{\frac{Q_P}{Q}} \,.
  \label{Eapprox}
\end{equation}

Using the conserved quantities, the other velocities are given by:
\begin{equation}
\frac{du }{d \tau} ~=~ \frac{E}{\sqrt{\cP} } \,(\cP \, Z_3 - V \, \mu^2)   \,, \qquad \frac{dv }{d \tau} ~=~  \frac{E}{\sqrt{\cP} } \,(\cP - K^3 \, \mu)    \,, \qquad \frac{d\psi }{d \tau} ~=~ \frac{E\, V\, \mu   }{\sqrt{\cP} }   \,.
  \label{othermotion}
\end{equation}
However, in the computation of the tidal forces it is simpler to use the frame velocities (\ref{framvels}).

\subsection{Tidal forces}
\label{ss:BRtidal}

Despite the simplifications, it is still a challenge to use Mathematica and the Riemann tensor in Appendix \ref{app:Curvatures}, to obtain the norm of the tidal tensor, $|\cA|$.  Indeed, it is simpler to work with $|\cA|^2$.  Because we are looking at geodesics on the $z$-axis for $z >\apoint$, this is purely a function of $z$  but it is the ratio of two degree-$48$ polynomials in $z$!

There are six control parameters of primary interest $\apoint, \bpoint, Q, Q_P, E$ and $z$, and we are particularly interested in the regime where
\begin{equation}
\apoint, \bpoint ~\ll~  Q, Q_P, z \,.
  \label{regime1}
\end{equation}
We therefore start by expanding in small $\apoint, \bpoint$.  The norm-square of the tidal tensor, $|\cA|^2$, is a quadratic in $E^2$, and the leading term in $E^4$ is
\begin{equation}
(\apoint- \bpoint)^2  \,E^4\,\bigg[  \frac{Q_P^2}{2\,z^6} \, (3 -4 \nu+ 2 \nu^2 )~+~  \frac{Q_P}{2\,z^5} \, (2 -5 \nu+ 4 \nu^2 )~+~  \frac{3}{16\,z^4} \, (1 -2 \nu)^2  \bigg]\,,\qquad \nu ~\equiv~ \frac{k_3^2 Q_P}{Q^2} \,.
  \label{E4sqleading}
\end{equation}
Indeed, at cubic order in $\apoint, \bpoint$, the $E^4$ term is proportional to $(\apoint -\bpoint)$.  

If one takes
\begin{equation}
\apoint, \bpoint ~\ll~  z ~\ll~ Q, Q_P \,, \qquad \nu \lesssim \cO(1)\,,
\label{regime2}
\end{equation}
then one arrives at
\begin{equation}
|\cA| ~\approx~ E^2\,  \frac{|\apoint- \bpoint|  \, Q_P}{|z|^3} \, \sqrt{3 -4 \nu+ 2 \nu^2}   \,.
  \label{E4leading}
\end{equation}
This is to be compared with (\ref{cAcappedBTZ}) and (\ref{asymmetry_tidal4}), and it is precisely the analogous term.  One should remember that in the regime (\ref{regime1}), on the $z$-axis, (\ref{bipolarchange})  reduces to  $z \approx \hat r \approx  \frac{1}{4} r^2$ and  $\apoint = \frac{1}{4}a^2$. 

Hence, this term may be rewritten, using  (\ref{JLJR}), as:
\begin{equation}
|\cA| ~\approx~ E^2\,  \frac{|J_L| \, Q_P}{8\, k_3\,|z|^3} \, \sqrt{3 -4 \nu+ 2 \nu^2} ~\approx~ E^2\,  \frac{8\, |J_L| \, Q_P}{ k_3\,r^6} \, \sqrt{3 -4 \nu+ 2 \nu^2}   \,.
  \label{E4leadingf}
\end{equation}

One can also look at other powers of $E^2$ in $|\cA|^2$ for the parameter range (\ref{regime2}).  While there are lower powers of $z^{-1}$,  the expansion does not have any net positive powers of $Q$ or $Q_P$ in the expansion, and so these tidal terms are sub-leading compared to  (\ref{E4sqleading}) especially when one takes into account the fact that $E$ is large.

Indeed, the only other interesting term in $|\cA|$ comes from setting $\apoint = \bpoint =0$:
\begin{equation}
|\cA|_{\apoint = \bpoint =0}  ~=~  \frac{1}{4 \sqrt{2}\, (Q+ 2 k \, k_3)} \,.
  \label{E4leadingAt0}
\end{equation}
In this limit the two black rings merge and form a black hole. As one can see from \eqref{ZFasymp}, the D1 and D5 five-dimensional supergravity charges of the solution, $Q_1$ and $Q_5$, are both equal to  $8 (Q+2 k k_3)$. Hence,  this term is exactly equal to the ``constant curvature'' tidal force felt by a particle falling into a black hole with charges   $Q_1$ and $Q_5$ (\ref{cAthroatBTZ}). 

If one sets $\apoint = \bpoint$, this sets $J_L=0$, and the  leading tidal terms described above vanish.  The leading term in $|\cA|^2$ at order $E^4$  now becomes
\begin{equation}
|\cA|^2 ~\approx~ E^4\,  \frac{18 \apoint^4 Q_P^2}{|z|^8} \,(3 -4 \nu+ 2 \nu^2)   \,.
  \label{E4sqleading2}
\end{equation}
or
\begin{equation}
|\cA| ~\approx~ E^2\,  \frac{3\sqrt{2} \, \apoint^2 \,Q_P}{|z|^4} \,\sqrt{3 -4 \nu+ 2 \nu^2} ~\approx~ E^2\,  \frac{48\sqrt{2} \, a^4 Q_P}{r^8} \,\sqrt{3 -4 \nu+ 2 \nu^2}  \,.
  \label{E4leading2}
\end{equation}

This term is the exact analog of (\ref{leadingAterm4}). This gives further confirmation to the fact that cancelling the global angular momentum, $J_L$, softens the tidal forces.

While the expression for the tidal tensor is generically extremely complicated, it is instructive to extract the terms that dominate at large $Q$ and then at large $Q_P$.  Specifically, as $Q \to \infty$, $\cA \sim Q^0$, and this term grows linearly in $Q_P$.  That is, for $Q \gg Q_p \gg
 z > a, b >0 $, we find:
\begin{equation}
|\cA| ~\approx~ E^2\, \sqrt{ \frac{3}{2}} \, Q_P \, \frac{(\apoint - \bpoint)\, z^3 + 6\, \apoint  \bpoint \,z^2 -3\,(\apoint - \bpoint)\, \apoint    \bpoint \,z +  \apoint  \bpoint\,( \apoint^2+ \bpoint^2)}{(z- \apoint)^3 (z+ \bpoint)^3}\,.
  \label{E4leading3}
\end{equation}
This reveals the leading-multipole expansion for the tidal force.  In particular, one sees the explicit role of the singular loci of the black rings. Moreover,  when $\apoint =\bpoint$, then the $\ZZ_2$ symmetry is restored and all the odd powers of $z$ disappear, leaving the $|z|^{-4}$ behavior of   (\ref{E4leading2}).  One also sees that,  in this limit, all the  higher, even multipoles of the tidal tensor are non-trivial.

\section{Final comments}
\label{sec:Conclusions}

As shown in \cite{Tyukov:2017uig,Bena:2018mpb} an observer falling into a  microstate geometry experiences a huge tidal disruption at a large distance away from the region where this geometry differs significantly from the black hole. The purpose of our investigation was to understand how much of this tidal disruption was caused by the finite angular momentum of the microstate geometries considered in \cite{Tyukov:2017uig,Bena:2018mpb}, and, more generally, how tidal forces  arise from the absence of spherical symmetry of the microstate geometry. 

We have seen that the leading-order contribution to the tidal disruption is proportional to the left-moving angular momentum and vanishes when the microstate geometry has a $\ZZ_2$ symmetry and hence no left-moving angular momentum. However, we found that the contribution at next order is finite, and does not vanish. Hence, an infalling observer in $\ZZ_2$-symmetric geometries still encounters a large tidal disruption away from the cap, but further down the throat than in geometries without the $Z_2$ symmetry.

One interesting question that merits further investigation is whether one can reduce the tidal disruption even more by considering even more specially-tuned microstate geometries in which the appropriate higher-order multi-pole contributions vanish as well.  It is evident from  (\ref{E4leading3}) that the tidal forces simply reflect the distribution of charge sources in the cap, and so we expect that higher multipoles could be cancelled by more finely-tuned cap structure.   Indeed, one might be able to reduce the tidal disruption of such microstate geometries to be of order
\begin{equation}
    \abs{\mathcal A} ~\sim~  \frac{  a^{2n} \, Q_P \,E^2 }{r^{4+2n}} \,, 
\label{lAterm-reduced}
\end{equation}
for some $n > 2$.  In this expression, $r$ is the radial coordinate in $\IR^4$ and $a$ is the scale of the cap.
 This would lead to Planckian tidal forces even closer to the horizon-sized structure, at a scale of order 
\begin{equation}
 r  ~\lesssim~    a^{1-\alpha} \, Q_P^{\frac{1}{2} \, \alpha} \,,   \qquad \alpha ~=~ \frac{3}{2(n+2)}    \,.
\label{rscale-reduced}
\end{equation}

It is also interesting to try to relate our tidal-disruption calculations to the recent calculation of gravitational multipoles of microstate geometries \cite{Bena:2020see}. Our paper has shown that the coefficient of the leading term in the tidal stress is given by the first angular momentum multipole. However, we have not identified the 
combination of multipoles that controls the next-to-leading-order term found in this paper \eqref{leadingAterm4},\eqref{E4leading2}, nor the next terms in the expansion conjectured in \eqref{lAterm-reduced}. It would be very interesting to identify whether the tidal-disruption terms we find are controlled by gravitational multipoles and, if so, what are the kinds of microstate geometries where the tidal disruption is the smallest. 

This would also allow us to understand whether a solution with a huge number of small bubbles, in an almost spherically symmetric configuration, could be arranged so that the tidal forces will remain very small until the infalling observer moves very close to the bubbles.   Indeed, if a typical solution involves a random distribution of a vast number, $N$, of bubbles, one would   expect that multipole moments would be suppressed by powers of $N$, and that one would only resolve the granularity of the bubbles, and be sensitive only higher multipole moments, when one is very close to the cap.

Another interesting question is whether the tidal-force calculation can offer us any hint on the type of black hole microstates that are dual to bubbling solutions. Indeed, unlike superstrata, for which the holographically-dual states are well understood \cite{Giusto:2015dfa,Galliani:2016cai,Ceplak:2018pws}, there is no known holographic dictionary for asymptotically-AdS$_3$ solutions with smoothly capped BTZ throats obtained from multiple bubbles. Furthermore, there are arguments that the states dual to multi-bubble solutions mix with other states as one moves away in moduli space \cite{Chowdhury:2013pqa, Bossard:2019ajg}.

On the other hand, we have seen that multi-bubble solutions give rise to tidal forces that can be much weaker than those of superstrata (in which the left-moving angular momentum never vanishes). Hence, an infalling observer in these geometries would, at least initially, experience a softer landing than in superstrata. This in turn might be argued, using fuzzball-complementarity philosophy \cite{Mathur:2012jk} to be a sign that the black-hole states described by multi-bubble microstate geometries with a long BTZ throat are closer to typicality than those constructed using present superstratum technology.

Another interesting question that our investigation opens is how much one expects the tidal forces felt by an observer in a typical microstate of a black hole to differ from those felt in the classical-gravity solution. The observation that tidal forces may be important at distances parametrically-larger than the size of the structure that replaces the black-hole horizon \cite{Tyukov:2017uig,Bena:2018mpb} gave one reasons to hope for a possible signature of this large tidal disruption in the gravitational waves emitted when two black holes merge. On the other hand, the result of this paper implies that in certain states the tidal disruption can be parametrically smaller than in others. Hence, we believe it is very important to understand whether the presence of large tidal disruptions far away from the microstucture is a feature of the typical states of the black hole or is an artifact of the atypicality of the microstate geometries that have been constructed in supergravity.


\section*{Acknowledgments}
\vspace{-2mm}
We would like to thank Daniel Mayerson for interesting discussion. The  work  of  IB is  supported  in part by  the  ANR  grant  Black-dS-String  ANR-16-CE31-0004-01, by the John Templeton Foundation grant 61149, and the ERC Grant 772408 -Stringlandscape. The work of NW is supported in part by the DOE grant DE-SC0011687. The work of IB, AH and NW is supported in part by the ERC Grant 787320 - QBH Structure. 

\appendix

\section{Six-dimensional connections and curvatures}
\label{app:Curvatures}

We start from the metric (\ref{sixmet2}) and the frames (\ref{sixframes}).

The frame connections, defined by $\dd{e}^a + \omega\indices{^a_b} \wedge e^b = 0$, are
\begin{align}
    \omega\indices{^0_1} ={}& -\frac 12 \stP^{-1/4} \hat{Z}_a e^{a+1}
    \\
    \omega\indices{^0_{a+1}} ={}& -\stP^{-1/4} \left(\frac 12 \hat{Z}_a + \frac 14 \hat{P}_a\right) e^0 - \frac 12 \hat{Z}_a \stP^{-1/4} e^1 + \frac 12 \stP^{-3/4} Z_3^{-1/2} (\dd{\omega}_{a b}) e^{b+1}
    \\
 \omega\indices{^1_{a+1}} ={}& \stP^{-1/4} \left(\frac 12 \hat{Z}_a - \frac 14 \hat{P}_a\right) e^1 + \frac 12 \hat{Z}_a \stP^{-1/4} e^0 + \frac 12 \stP^{-3/4}  \left(Z_3^{1/2} \dd{\beta}_{ab} - Z_3^{-1/2} \dd{\omega}_{ab} \right) (\dd{\omega}_{a b}) e^{b+1}
    \\
    \begin{split}
        \omega\indices{^{a+1}_{b+1}} ={}& \stP^{-1/4} \hat{\omega}\indices{_d^a_b} e^{d+1} + \frac 14 \stP^{-1/4} \qty( \hat{P}_b e^{a+1} - \hat{P}_a e^{b+1} ) + \frac 12 Z_3^{-1/2} \stP^{-3/4} (\dd{\omega}_{ab}) e^0
        \\
        &- \frac 12 \stP^{-3/4} \qty(Z_3^{1/2} \dd{\beta}_{ab} - Z_3^{-1/2} \dd{\omega}_{ab}) e^1
    \end{split}
\end{align}
where  $ \hat{\omega}\indices{_d^a_b} \hat e^{d+1}$ is the spin connection on the four-dimensional base manifold, and where
\begin{align}
    \hat{P}_b & \equiv \left( \stP^{-1} \partial_\mu \stP \right) \hat{e}\indices{^\mu_b}
    \\
    \hat{Z}_b &\equiv \left( Z_3^{-1} \partial_\mu Z_3 \right) \hat{e}\indices{^\mu_b}
    \\
    \dd{\omega}_{ab} &\equiv \qty(\partial_\mu \omega_\nu - \partial_\nu \omega_\mu) \hat{e}\indices{^\mu_a} \wedge \hat{e}\indices{^\nu_b}
    \\
    \dd{\beta}_{ab} &\equiv \qty(\partial_\mu \beta_\nu - \partial_\nu \beta_\mu) \hat{e}\indices{^\mu_a} \wedge \hat{e}\indices{^\nu_b}
\end{align}
The components of the 2-form, $\dd{\omega}_{ab}$, should not be confused with the spin connection. 

The Riemann curvature tensor 2-form can then be computed through the formula $R\indices{^a_b} = \dd{\omega}\indices{^a_b} + \omega\indices{^a_c}\wedge \omega\indices{^c_b}$. The result is :

\begin{equation}
\begin{split}
    R\indices{^0_1} ={}& -\frac 1{16} \stP^{-1/2}   \qty(\hat{P}_a\hat{P}_a) e^0 \wedge e^1
    \\
    & - \frac 14 \stP^{-1} \qty( \frac 12 Z_3^{-1/2} \hat{P}_a \dd{\omega}_{ab} - Z_3^{1/2} \qty(\hat{Z}_a + \frac 12 \hat{P}_a) \dd{\beta}_{ab}) e^0 \wedge e^{b+1}
    \\
    & - \frac 14 \stP^{-1} \qty( \frac 12 Z_3^{-1/2} \hat{P}_a \dd{\omega}_{ab} - Z_3 ^{1/2} \hat{Z_a}\dd{\beta}_{ab}) e^1 \wedge e^{b+1}
    \\
    & - \frac 14 \stP^{-3/2} \dd{\omega}_{ca}\dd{\beta}_{cb} e^{a+1} \wedge e^{b+1}
\end{split}
\end{equation}

\begin{equation}
\begin{split}
    R\indices{^0_{a+1}} &= \stP^{-1/2} \bigg( -\frac 3{16} \hat{P}_a \hat{P}_b - \frac 14 \hat{Z}_a \hat{P}_b - \frac 14 \hat{Z}_b \hat{P}_a + \frac 12 \hat{Z}_a \hat{Z}_b  + \frac 12 \hat{\nabla}_b \hat{Z}_a + \frac 14 \hat{\nabla}_b \hat{\stP}_a \\ 
    & \qquad\qquad\quad - \frac 14 Z_3^{-1} \stP^{-1} \dd{\omega}_{ca} \dd{\omega}_{cb}\bigg) e^0 \wedge e^{b+1}
    \\
    & + \stP^{-1/2} \qty( \frac 18 \hat{Z}_b \hat{P}_b + \frac 1{16} \hat{P}_b \hat{P}_b) e^0 \wedge e^{a+1} + \frac 18 \stP^{-1/2} \qty(\hat{Z}_b \hat{P}_b) e^1 \wedge e^{a+1} 
    \\
    & + \stP^{-1/2} \bigg(- \frac 14 \hat{Z}_a \hat{P}_b - \frac 14 \hat{Z}_b \hat{P}_a + \frac 12 \hat{Z}_a \hat{Z}_b + \frac 12 \hat{\nabla}_b \hat{Z}_a \\
    & \qquad\qquad + \frac 14 \stP^{-1} \dd{\omega}_{cb} \qty(\dd{\beta}_{ca} - Z_3^{-1} \dd{\omega}_{ca})\bigg) e^1 \wedge e^{b+1}
    \\
    & + \stP^{-1} Z_3^{1/2} \qty(\frac 14 \qty(\hat{Z}_b + \frac 12 \hat{P}_b) \dd{\beta}_{ba} - \frac 18 Z_3^{-1} \hat{P}_b \dd{\omega}_{ba}) e^0 \wedge e^1
    \\
    & + \frac 18 \stP^{-1} Z_3^{-1/2}  \hat{P}_c \dd{\omega}_{cb} e^{a+1} \wedge e^{b+1}
    \\
    & + \stP^{-1} Z_3^{1/2} \bigg( - \frac 14 Z_3^{-1} \hat{P}_a \dd{\omega}_{bc} - \frac 14 \hat{Z}_a \dd{\beta}_{bc} + \frac 14 Z_3^{-1} \hat{P}_c \dd{\omega}_{ab} - \frac 14 \hat{Z}_b \dd{\beta}_{ac} \\
    & \qquad\qquad\qquad  - \frac 12 Z_3^{-1} \hat{\nabla}_c (\dd{\omega}_{ab})\bigg) e^{b+1} \wedge e^{c+1}
\end{split}
\end{equation}

\begin{equation}
\begin{split}
    R\indices{^1_{a+1}} &= \stP^{-1/2} \qty( \frac 14 \qty(\hat{Z}_a \hat{P}_b + \hat{Z}_b \hat{P}_a) - \frac 12 \hat{Z}_a \hat{Z}_b - \frac 12 \hat{\nabla}_b \hat{Z}_a + \frac 14 \stP^{-1} \dd{\omega}_{ac} \qty(\dd{\beta}_{cb} - Z_3^{-1} \dd{\omega}_{cb}) ) e^0 \wedge e^{b+1}
    \\
    & - \frac 18 \stP^{-1/2} \qty(  \hat{Z}_b \hat{P}_b) e^0 \wedge e^{a+1}
    \\
    & \begin{split}
         + \stP^{-1/2} \Biggl(& \frac 14 \qty(\hat{Z}_a \hat{P}_b + \hat{Z}_b \hat{P}_a) - \frac 3{16} \hat{P}_a \hat{P}_b - \frac 12 \hat{Z}_a \hat{Z}_b - \frac 12 \hat{\nabla}_b \hat{Z}_a + \frac 14 \hat{\nabla}_b \hat{P}_a
         \\
        &  \qquad\qquad - \frac 14 \stP^{-1} Z_3 \qty(\dd{\beta}_{ac} - Z_3^{-1} \dd{\omega}_{ac})\qty(\dd{\beta}_{cb} - Z_3^{-1} \dd{\omega}_{cb})\Biggr) e^1 \wedge e^{b+1}
    \end{split}
    \\
    & - \stP^{-1/2} \qty(\frac 18 \hat{Z}_b \hat{P}_b - \frac 1{16} \hat{P}_b \hat{P}_b) e^1 \wedge e^{a+1}
    \\
    & + \stP^{-1} Z_3^{1/2} \qty(\frac 18 \hat{P}_b Z_3^{-1} \dd{\omega}_{ba} - \frac 14 \hat{Z}_b \dd{\beta}_{ba}) e^0 \wedge e^1
    \\
    & +  \frac 18  \stP^{-1} Z_3^{1/2}\hat{P}_c \qty(\dd{\beta}_{cb} - Z_3^{-1} \dd{\omega}_{cb}) e^{a+1} \wedge e^{b+1}
    \\
    & \begin{split}
         + \stP^{-1} Z_3^{1/2} \Biggl( & \frac 14 Z_3^{-1} \hat{P}_a \dd{\omega}_{bc} + \frac 14 \qty(\hat{Z}_a - \hat{P}_a) \dd{\beta}_{bc} + \frac 14 Z_3^{-1} \hat{P}_c \dd{\omega}_{ab} + \frac 14 \qty(\hat{P}_c - \hat{Z}_c) \dd{\beta}_{ab}
         \\
         & \qquad\qquad  - \frac 12 \hat{\nabla}_c (\dd{\beta}_{ab}) + \frac 12 Z_3^{-1} \hat{\nabla}_c (\dd{\omega}_{ab})\Biggr) e^{b+1} \wedge e^{c+1}
    \end{split}
\end{split}
\end{equation}

\begin{equation}
\begin{split}
    R\indices{^{a+1}_{b+1}} &= \stP^{-1/2} \hat{R}\indices{^a_b}
    \\
    & - \frac14 \stP^{-3/2} \qty( \dd{\omega}_{ae} \dd{\beta}_{eb} - \dd{\omega}_{be} \dd{\beta}_{ea} ) e^0 \wedge e^1
    \\
    & \begin{split}
        + \stP^{-1} Z_3^{1/2} \Biggl(& \frac 12 \hat{P}_c Z_3^{-1} \dd{\omega}_{ab} + \frac 14 Z_3^{-1} \hat{P}_b \dd{\omega}_{ac} - \frac 14 Z_3^{-1} \hat{P}_a \dd{\omega}_{bc} - \frac12 Z_3^{-1} \hat{\nabla}_c \dd{\omega}_{ab}
        \\
        & \qquad \qquad+ \frac12 \hat{Z_c} \dd{\beta}_{ab} + \frac14 \hat{Z}_b \dd{\beta}_{ac} - \frac14 \hat{Z}_a \dd{\beta}_{bc} \Biggr) e^0 \wedge e^{c+1}
    \end{split}
    \\
    & \begin{split}
        + \stP^{-1} Z_3^{1/2} \Biggl(& \frac 12 \qty(\hat{Z}_c - \hat{P}_c) \dd{\beta}_{ab} + \frac 12 Z_3^{-1} \hat{P}_c \dd{\omega}_{ab} + \frac12  \hat{\nabla}_c (\dd{\beta}_{ab})  - \frac12 Z_3^{-1} \hat{\nabla}_c (\dd{\omega}_{ab})
        \\
        & - \frac14 Z_3^{-1} \qty(\hat{P}_a \dd{\omega}_{bc} - \hat{P}_b \dd{\omega}_{ac}) \\
        & \qquad \qquad\qquad\qquad +  \frac14 \Big( (\hat{P}_a - \hat{Z}_a) \dd{\beta}_{bc} - (\hat{P}_b-\hat{Z}_b) \dd{\beta}_{ac}\Big) \Biggr) e^1 \wedge e^{c+1}
    \end{split}
    \\
    & - \frac18 \stP^{-1} Z_3^{-1/2} \qty(\hat{P}_c \dd{\omega}_{cb} e^0 \wedge e^{a+1} - \hat{P}_c \dd{\omega}_{ca} e^0 \wedge e^{b+1})
    \\
    & + \frac18 \stP^{-1} Z_3^{1/2} \hat{P}_c \qty( \qty(\dd{\beta}_{cb} - Z_3^{-1}\dd{\omega}_{cb}) e^1 \wedge e^{a+1} - \qty(\dd{\beta}_{ca} - Z_3^{-1}\dd{\omega}_{ca}) e^1 \wedge e^{b+1})
    \\
    & \begin{split}
        + \frac14 \stP^{-3/2} \Bigl(&- Z_3 \dd{\beta}_{ab} \dd{\beta}_{cd} + \dd{\beta}_{ab}\dd{\omega}_{cd} + \dd{\beta}_{cd}\dd{\omega}_{ab}
        \\
        & \qquad \qquad\qquad\qquad  - Z_3 \dd{\beta}_{ac}\dd{\beta}_{bd} + \dd{\beta}_{ac}\dd{\omega}_{bd} + \dd{\beta}_{bd}\dd{\omega}_{ac}\Bigr) e^{c+1} \wedge e^{d+1}
    \end{split}
    \\
    & + \stP^{-1/2} \qty(\frac1{16} \hat{P}_b \hat{P}_c - \frac14 \hat{\nabla}_c\hat{P}_b) e^{a+1} \wedge e^{c+1}
     - \stP^{-1/2} \qty(\frac1{16} \hat{P}_a \hat{P}_c - \frac14 \hat{\nabla}_c\hat{P}_a) e^{b+1} \wedge e^{c+1}
    \\
    & - \frac1{16} \stP^{-1/2} \qty(\hat{P}_c \hat{P}_c) e^{a+1} \wedge e^{b+1}
    \\
\end{split}
\end{equation}

We have used the curvature tensor of the base space $\hat{R}\indices{^a_b}$, and its covariant derivative $\hat{\nabla}$.


\begin{adjustwidth}{-1mm}{-1mm} 

\bibliographystyle{utphys}

\bibliography{microstates}       

\providecommand{\href}[2]{#2}\begingroup\raggedright\begin{thebibliography}{10}

\bibitem{Bena:2006kb}
I.~Bena, C.-W. Wang, and N.~P. Warner, ``{Mergers and Typical Black Hole
  Microstates},'' \href{http://dx.doi.org/10.1088/1126-6708/2006/11/042}{{\em
  JHEP} {\bfseries 11} (2006) 042},
\href{http://arxiv.org/abs/hep-th/0608217}{{\ttfamily arXiv:hep-th/0608217}}.

\bibitem{Bena:2007qc}
I.~Bena, C.-W. Wang, and N.~P. Warner, ``{Plumbing the Abyss: Black Ring
  Microstates},'' \href{http://dx.doi.org/10.1088/1126-6708/2008/07/019}{{\em
  JHEP} {\bfseries 07} (2008) 019},
\href{http://arxiv.org/abs/0706.3786}{{\ttfamily arXiv:0706.3786 [hep-th]}}.

\bibitem{Bena:2016ypk}
I.~Bena, S.~Giusto, E.~J. Martinec, R.~Russo, M.~Shigemori, D.~Turton, and
  N.~P. Warner, ``{Smooth horizonless geometries deep inside the black-hole
  regime},'' \href{http://dx.doi.org/10.1103/PhysRevLett.117.201601}{{\em Phys.
  Rev. Lett.} {\bfseries 117} no.~20, (2016) 201601},
\href{http://arxiv.org/abs/1607.03908}{{\ttfamily arXiv:1607.03908 [hep-th]}}.

\bibitem{Bena:2017xbt}
I.~Bena, S.~Giusto, E.~J. Martinec, R.~Russo, M.~Shigemori, D.~Turton, and
  N.~P. Warner, ``{Asymptotically-flat supergravity solutions deep inside the
  black-hole regime},'' \href{http://dx.doi.org/10.1007/JHEP02(2018)014}{{\em
  JHEP} {\bfseries 02} (2018) 014},
\href{http://arxiv.org/abs/1711.10474}{{\ttfamily arXiv:1711.10474 [hep-th]}}.

\bibitem{Heidmann:2017cxt}
P.~Heidmann, ``{Four-center bubbled BPS solutions with a Gibbons-Hawking
  base},'' \href{http://dx.doi.org/10.1007/JHEP10(2017)009}{{\em JHEP}
  {\bfseries 10} (2017) 009},
\href{http://arxiv.org/abs/1703.10095}{{\ttfamily arXiv:1703.10095 [hep-th]}}.

\bibitem{Bena:2017fvm}
I.~Bena, P.~Heidmann, and P.~F. Ramirez, ``{A systematic construction of
  microstate geometries with low angular momentum},''
  \href{http://dx.doi.org/10.1007/JHEP10(2017)217}{{\em JHEP} {\bfseries 10}
  (2017) 217},
\href{http://arxiv.org/abs/1709.02812}{{\ttfamily arXiv:1709.02812 [hep-th]}}.

\bibitem{Avila:2017pwi}
J.~Avila, P.~F. Ramirez, and A.~Ruiperez, ``{One Thousand and One Bubbles},''
  \href{http://dx.doi.org/10.1007/JHEP01(2018)041}{{\em JHEP} {\bfseries 01}
  (2018) 041},
\href{http://arxiv.org/abs/1709.03985}{{\ttfamily arXiv:1709.03985 [hep-th]}}.

\bibitem{Ceplak:2018pws}
N.~{\v C}eplak, R.~Russo, and M.~Shigemori, ``{Supercharging Superstrata},''
  \href{http://dx.doi.org/10.1007/JHEP03(2019)095}{{\em JHEP} {\bfseries 03}
  (2019) 095}, \href{http://arxiv.org/abs/1812.08761}{{\ttfamily
  arXiv:1812.08761 [hep-th]}}.

\bibitem{Heidmann:2019zws}
P.~Heidmann and N.~P. Warner, ``{Superstratum Symbiosis},''
  \href{http://dx.doi.org/10.1007/JHEP09(2019)059}{{\em JHEP} {\bfseries 09}
  (2019) 059}, \href{http://arxiv.org/abs/1903.07631}{{\ttfamily
  arXiv:1903.07631 [hep-th]}}.

\bibitem{Heidmann:2019xrd}
P.~Heidmann, D.~R. Mayerson, R.~Walker, and N.~P. Warner, ``{Holomorphic Waves
  of Black Hole Microstructure},''
  \href{http://dx.doi.org/10.1007/JHEP02(2020)192}{{\em JHEP} {\bfseries 02}
  (2020) 192}, \href{http://arxiv.org/abs/1910.10714}{{\ttfamily
  arXiv:1910.10714 [hep-th]}}.

\bibitem{Warner:2019jll}
N.~P. Warner, ``{Lectures on Microstate Geometries},''
  \href{http://arxiv.org/abs/1912.13108}{{\ttfamily arXiv:1912.13108
  [hep-th]}}.

\bibitem{Mathur:2009hf}
S.~D. Mathur, ``{The information paradox: A pedagogical introduction},''
  \href{http://dx.doi.org/10.1088/0264-9381/26/22/224001}{{\em Class. Quant.
  Grav.} {\bfseries 26} (2009) 224001},
\href{http://arxiv.org/abs/0909.1038}{{\ttfamily arXiv:0909.1038 [hep-th]}}.

\bibitem{Almheiri:2012rt}
A.~Almheiri, D.~Marolf, J.~Polchinski, and J.~Sully, ``{Black Holes:
  Complementarity or Firewalls?},''
  \href{http://dx.doi.org/10.1007/JHEP02(2013)062}{{\em JHEP} {\bfseries 1302}
  (2013) 062},
\href{http://arxiv.org/abs/1207.3123}{{\ttfamily arXiv:1207.3123 [hep-th]}}.

\bibitem{Kraus:2015zda}
P.~Kraus and S.~D. Mathur, ``{Nature abhors a horizon},''
  \href{http://dx.doi.org/10.1142/S0218271815430038}{{\em Int. J. Mod. Phys. D}
  {\bfseries 24} no.~12, (2015) 1543003},
  \href{http://arxiv.org/abs/1505.05078}{{\ttfamily arXiv:1505.05078
  [hep-th]}}.

\bibitem{Bena:2015dpt}
I.~Bena, D.~R. Mayerson, A.~Puhm, and B.~Vercnocke, ``{Tunneling into
  Microstate Geometries: Quantum Effects Stop Gravitational Collapse},''
  \href{http://dx.doi.org/10.1007/JHEP07(2016)031}{{\em JHEP} {\bfseries 07}
  (2016) 031}, \href{http://arxiv.org/abs/1512.05376}{{\ttfamily
  arXiv:1512.05376 [hep-th]}}.

\bibitem{Tyukov:2017uig}
A.~Tyukov, R.~Walker, and N.~P. Warner, ``{Tidal Stresses and Energy Gaps in
  Microstate Geometries},''
  \href{http://dx.doi.org/10.1007/JHEP02(2018)122}{{\em JHEP} {\bfseries 02}
  (2018) 122},
\href{http://arxiv.org/abs/1710.09006}{{\ttfamily arXiv:1710.09006 [hep-th]}}.

\bibitem{Bena:2018mpb}
I.~Bena, E.~J. Martinec, R.~Walker, and N.~P. Warner, ``{Early Scrambling and
  Capped BTZ Geometries},''
  \href{http://dx.doi.org/10.1007/JHEP04(2019)126}{{\em JHEP} {\bfseries 04}
  (2019) 126},
\href{http://arxiv.org/abs/1812.05110}{{\ttfamily arXiv:1812.05110 [hep-th]}}.

\bibitem{Giusto:2015dfa}
S.~Giusto, E.~Moscato, and R.~Russo, ``{AdS$_{3}$ holography for 1/4 and 1/8
  BPS geometries},'' \href{http://dx.doi.org/10.1007/JHEP11(2015)004}{{\em
  JHEP} {\bfseries 11} (2015) 004},
\href{http://arxiv.org/abs/1507.00945}{{\ttfamily arXiv:1507.00945 [hep-th]}}.

\bibitem{Galliani:2016cai}
A.~Galliani, S.~Giusto, E.~Moscato, and R.~Russo, ``{Correlators at large c
  without information loss},''
  \href{http://dx.doi.org/10.1007/JHEP09(2016)065}{{\em JHEP} {\bfseries 09}
  (2016) 065}, \href{http://arxiv.org/abs/1606.01119}{{\ttfamily
  arXiv:1606.01119 [hep-th]}}.

\bibitem{Bena:2018bbd}
I.~Bena, P.~Heidmann, and D.~Turton, ``{AdS$_{2}$ holography: mind the cap},''
  \href{http://dx.doi.org/10.1007/JHEP12(2018)028}{{\em JHEP} {\bfseries 12}
  (2018) 028},
\href{http://arxiv.org/abs/1806.02834}{{\ttfamily arXiv:1806.02834 [hep-th]}}.

\bibitem{Bena:2015bea}
I.~Bena, S.~Giusto, R.~Russo, M.~Shigemori, and N.~P. Warner, ``{Habemus
  Superstratum! A constructive proof of the existence of superstrata},''
  \href{http://dx.doi.org/10.1007/JHEP05(2015)110}{{\em JHEP} {\bfseries 05}
  (2015) 110},
\href{http://arxiv.org/abs/1503.01463}{{\ttfamily arXiv:1503.01463 [hep-th]}}.

\bibitem{Gutowski:2003rg}
J.~B. Gutowski, D.~Martelli, and H.~S. Reall, ``{All supersymmetric solutions
  of minimal supergravity in six dimensions},''
  \href{http://dx.doi.org/10.1088/0264-9381/20/23/008}{{\em Class. Quant.
  Grav.} {\bfseries 20} (2003) 5049--5078},
\href{http://arxiv.org/abs/hep-th/0306235}{{\ttfamily arXiv:hep-th/0306235}}.

\bibitem{Peet:2000hn}
A.~W. Peet, ``{TASI lectures on black holes in string theory},''
\href{http://arxiv.org/abs/hep-th/0008241}{{\ttfamily arXiv:hep-th/0008241}}.

\bibitem{Gauntlett:2004qy}
J.~P. Gauntlett and J.~B. Gutowski, ``{General concentric black rings},''
  \href{http://dx.doi.org/10.1103/PhysRevD.71.045002}{{\em Phys.Rev.}
  {\bfseries D71} (2005) 045002},
\href{http://arxiv.org/abs/hep-th/0408122}{{\ttfamily arXiv:hep-th/0408122
  [hep-th]}}.

\bibitem{Bena:2007kg}
I.~Bena and N.~P. Warner, ``{Black holes, black rings and their microstates},''
  \href{http://dx.doi.org/10.1007/978-3-540-79523-0}{{\em Lect. Notes Phys.}
  {\bfseries 755} (2008) 1--92},
\href{http://arxiv.org/abs/hep-th/0701216}{{\ttfamily arXiv:hep-th/0701216}}.

\bibitem{Bena:2005va}
I.~Bena and N.~P. Warner, ``{Bubbling supertubes and foaming black holes},''
  \href{http://dx.doi.org/10.1103/PhysRevD.74.066001}{{\em Phys. Rev.}
  {\bfseries D74} (2006) 066001},
\href{http://arxiv.org/abs/hep-th/0505166}{{\ttfamily arXiv:hep-th/0505166}}.

\bibitem{Berglund:2005vb}
P.~Berglund, E.~G. Gimon, and T.~S. Levi, ``{Supergravity microstates for BPS
  black holes and black rings},''
  \href{http://dx.doi.org/10.1088/1126-6708/2006/06/007}{{\em JHEP} {\bfseries
  0606} (2006) 007},
\href{http://arxiv.org/abs/hep-th/0505167}{{\ttfamily arXiv:hep-th/0505167
  [hep-th]}}.

\bibitem{Bates:2003vx}
B.~Bates and F.~Denef, ``{Exact solutions for supersymmetric stationary black
  hole composites},'' \href{http://dx.doi.org/10.1007/JHEP11(2011)127}{{\em
  JHEP} {\bfseries 1111} (2011) 127},
\href{http://arxiv.org/abs/hep-th/0304094}{{\ttfamily arXiv:hep-th/0304094
  [hep-th]}}.

\bibitem{Gauntlett:2004wh}
J.~P. Gauntlett and J.~B. Gutowski, ``{Concentric black rings},''
  \href{http://dx.doi.org/10.1103/PhysRevD.71.025013}{{\em Phys. Rev. D}
  {\bfseries 71} (2005) 025013},
  \href{http://arxiv.org/abs/hep-th/0408010}{{\ttfamily arXiv:hep-th/0408010}}.

\bibitem{Chowdhury:2013pqa}
B.~D. Chowdhury and D.~R. Mayerson, ``{Multi-centered D1-D5 solutions at finite
  B-moduli},'' \href{http://dx.doi.org/10.1007/JHEP02(2014)043}{{\em JHEP}
  {\bfseries 02} (2014) 043},
\href{http://arxiv.org/abs/1305.0831}{{\ttfamily arXiv:1305.0831 [hep-th]}}.

\bibitem{Bena:2004tk}
I.~Bena and P.~Kraus, ``{Microscopic description of black rings in AdS/CFT},''
  \href{http://dx.doi.org/10.1088/1126-6708/2004/12/070}{{\em JHEP} {\bfseries
  12} (2004) 070},
\href{http://arxiv.org/abs/hep-th/0408186}{{\ttfamily arXiv:hep-th/0408186}}.

\bibitem{Bena:2017geu}
I.~Bena, E.~Martinec, D.~Turton, and N.~P. Warner, ``{M-theory Superstrata and
  the MSW String},'' \href{http://dx.doi.org/10.1007/JHEP06(2017)137}{{\em
  JHEP} {\bfseries 06} (2017) 137},
\href{http://arxiv.org/abs/1703.10171}{{\ttfamily arXiv:1703.10171 [hep-th]}}.

\bibitem{Bena:2016agb}
I.~Bena, E.~Martinec, D.~Turton, and N.~P. Warner, ``{Momentum Fractionation on
  Superstrata},'' \href{http://dx.doi.org/10.1007/JHEP05(2016)064}{{\em JHEP}
  {\bfseries 05} (2016) 064},
\href{http://arxiv.org/abs/1601.05805}{{\ttfamily arXiv:1601.05805 [hep-th]}}.

\bibitem{Bena:2008wt}
I.~Bena, N.~Bobev, and N.~P. Warner, ``{Spectral Flow, and the Spectrum of
  Multi-Center Solutions},''
  \href{http://dx.doi.org/10.1103/PhysRevD.77.125025}{{\em Phys. Rev.}
  {\bfseries D77} (2008) 125025},
\href{http://arxiv.org/abs/0803.1203}{{\ttfamily arXiv:0803.1203 [hep-th]}}.

\bibitem{Bena:2020see}
I.~Bena and D.~R. Mayerson, ``{A New Window into Black Holes},''
  \href{http://arxiv.org/abs/2006.10750}{{\ttfamily arXiv:2006.10750
  [hep-th]}}.

\bibitem{Bossard:2019ajg}
G.~Bossard and S.~Luest, ``{Microstate geometries at a generic point in moduli
  space},'' \href{http://dx.doi.org/10.1007/s10714-019-2584-4}{{\em Gen. Rel.
  Grav.} {\bfseries 51} no.~9, (2019) 112},
  \href{http://arxiv.org/abs/1905.12012}{{\ttfamily arXiv:1905.12012
  [hep-th]}}.

\bibitem{Mathur:2012jk}
S.~D. Mathur and D.~Turton, ``{Comments on black holes I: The possibility of
  complementarity},'' \href{http://dx.doi.org/10.1007/JHEP01(2014)034}{{\em
  JHEP} {\bfseries 1401} (2014) 034},
\href{http://arxiv.org/abs/1208.2005}{{\ttfamily arXiv:1208.2005 [hep-th]}}.

\end{thebibliography}\endgroup

\end{adjustwidth}

\end{document}